\documentclass{article}
\usepackage[super,sort&compress,comma]{natbib} 
\usepackage{balance}
\usepackage{mathptmx}
\usepackage{graphicx} 
\usepackage{float}
\usepackage{hyperref}

\usepackage{textgreek}
\usepackage{booktabs}
\usepackage{multirow}

\usepackage{amsmath,amsfonts,bm}
\newcommand{\Var}{\mathrm{Var}}
\newcommand{\E}{\mathbb{E}}

\title{Graph Neural Network Interatomic Potential Ensembles with Calibrated Aleatoric and Epistemic Uncertainty on Energy and Forces$^\dag$}

\author{
    Jonas Busk\textit{$^{a,\ast}$} \and 
    Mikkel N. Schmidt\textit{$^{b}$} \and 
    Ole Winther{$^{b,c,d}$} \and 
    Tejs Vegge\textit{$^{a}$} \and 
    Peter Bj{\o}rn J{\o}rgensen\textit{$^{a}$}
}

\date{\small{
    $^{a}$Department of Energy Conversion and Storage, Technical University of Denmark, Kongens Lyngby, Denmark. E-mail: \{jbusk,teve,pbjo\}@dtu.dk. \\
    $^{b}$Department of Applied Mathematics and Computer Science, Technical University of Denmark, Kongens Lyngby, Denmark. E-mail: \{mnsc,olwi\}@dtu.dk. \\
    $^{c}$Center for Genomic Medicine, Rigshospitalet, Copenhagen University Hospital, Denmark. \\
    $^{d}$Bioinformatics Centre, Department of Biology, University of Copenhagen, Denmark. \\
    $^{*}$Corresponding author: jbusk@dtu.dk.\\
    $^\dag$Electronic Supplementary Information (ESI) available.
}}

\begin{document}

\maketitle

\begin{abstract}
Inexpensive machine learning potentials are increasingly being used to speed up structural optimization and molecular dynamics simulations of materials by iteratively predicting and applying interatomic forces.
In these settings, it is crucial to detect when predictions are unreliable to avoid wrong or misleading results.
Here, we present a complete framework for training and recalibrating graph neural network ensemble models to produce accurate predictions of energy and forces with calibrated uncertainty estimates.
The proposed method considers both epistemic and aleatoric uncertainty and the total uncertainties are recalibrated post hoc using a nonlinear scaling function to achieve good calibration on previously unseen data, without loss of predictive accuracy.
The method is demonstrated and evaluated on two challenging, publicly available datasets, ANI-1x~(\citet{ani1x}) and Transition1x~(\citet{schreiner2022}), both containing diverse conformations far from equilibrium. 
A detailed analysis of the predictive performance and uncertainty calibration is provided.
In all experiments, the proposed method achieved low prediction error and good uncertainty calibration, with predicted uncertainty correlating with expected error, on energy and forces.
To the best of our knowledge, the method presented in this paper is the first to consider a complete framework for obtaining calibrated epistemic and aleatoric uncertainty predictions on both energy and forces in ML potentials.
\end{abstract}

\section{Introduction}

Accurate and computationally inexpensive machine learning (ML) potentials are increasingly being used to accelerate atomic structure optimization and molecular dynamics simulations by iteratively predicting and applying interatomic energies and forces~\citep{dral2020, lilienfeld2020}.
This development has the potential to revolutionise disciplines of computational chemistry such as predicting molecular properties and structures, predicting reaction mechanisms and networks, as well as discovering new materials, e.g., for energy conversion and storage of renewable energy.
In these settings, it is crucial to asses the confidence of predictions and to detect when predictions are unreliable, to avoid wrong or misleading results by either ending the simulation early or enabling recovery by falling back to higher fidelity but also more computationally expensive methods, such as density functional theory (DFT)~\citep{peterson2017}.
Uncertainty quantification (UQ) methods can enable assessment of the confidence in predictions and thus make applications of ML potentials more robust and reliable.
To ensure uncertainty estimates are useful and informative, they need to be calibrated, i.e. there should be an agreement between the predictive distribution and the empirical distribution.
Especially if the predicted uncertainty is expected to indicate the range of plausible values.
For example in a screening application where candidate materials are filtered for specific useful properties, instances with poor point estimates but with high uncertainty could still potentially be interesting and should not be discarded.
Good calibration thus ensures that ML-based uncertainty estimates are interpretable and actionable, and enable the selection of a suitable confidence threshold for a given application on the original unit scale of the quantity of interest.

When considering predictive uncertainty it is often useful to distinguish between \emph{epistemic uncertainty} and \emph{aleatoric uncertainty}~\citep{uncertainty_in_ml, gal2017}.
Epistemic uncertainty arises from uncertainty in determining the model parameters and can in principle be reduced by observing more data.
On the other hand, aleatoric uncertainty can originate from random noise or inconsistency in the data, or from inadequacy of the model to fit the data precisely, and can therefore generally not be reduced by observing more data.
A widely used method for estimating epistemic uncertainty in ML potentials is to apply an ensemble of models and use the agreement of the predictions of the ensemble members as a measure of confidence in the prediction~\citep{schran2020}.
This approach relies on the observation that randomly initialised models will often provide increasingly different predictions further away from the training data distribution.
From a Bayesian perspective, if the individual ensemble members are seen as draws from the posterior distribution, the variance between predictions of the ensemble members are a measure of the posterior uncertainty~\citep{wilson2020, hoffmann2021, gustaffson2020}.
Other popular approaches for estimating epistemic uncertainty with neural networks include Bayesian neural networks~\cite{neal1993, Blundell2015} that can directly learn a probability distribution over the neural network parameters, and Monte Carlo dropout~\citep{gal2016} that estimates the predictive distribution through multiple stochastic forward passes.
However, UQ methods that only account for epistemic uncertainty and thus ignore aleatoric uncertainty are not inherently calibrated, which makes it difficult to select an appropriate confidence threshold for a given application.
Methods for estimating the aleatoric uncertainty include the mean-variance model~\citep{nix}, that explicitly predicts the uncertainty variance as an additional model output, and more recently evidential learning~\citep{amini2020, soleimany2021} that learns the parameters of a higher order distribution over the likelihood parameters, and conformal prediction~\citep{hu2022}, a distribution-free approach that estimates a prediction interval directly.
The deep ensemble approach~\citep{deepensembles} combines an ensemble of mean-variance neural network models to estimate both aleatoric and epistemic uncertainty in a unified model.

As discussed above, an important aspect of predictive uncertainty is the concept of \emph{calibration}, which implies an agreement between the predicted uncertainty and the expected empirical error.
When evaluating calibration, it is important to consider the asymmetric relationship between errors and uncertainties.
By the common assumption that errors are drawn from a distribution (usually Gaussian), small uncertainties should be associated with small errors and large errors should be associated with large uncertainties, but large uncertainties can be associated with both small and large errors.
Therefore there is no direct correlation between uncertainties and the magnitude of errors.
However, there should be a correlation between the uncertainties and the expected magnitude of errors.
Several works have proposed methods for evaluating and validation calibration of regression models. 
\citet{kuleshov2018} proposed evaluating the coverage of the errors by the predictive distribution averaged over the data using a calibration curve.
Later, \citet{levi2022} proposed checking the correlation of uncertainties and expected errors computed in bins of increasing uncertainty in a reliability diagram.
Recently, \citet{pernot2022} highlighted the limitations of the previous approaches and proposed an additional analysis of z-scores (standard scores) for variance-based UQ methods.
UQ and calibration for molecular property prediction and interatomic ML potentials has been explored in recent literature~\citep{tran2020, hirschfeld2020, scalia2020, nigam2021, hu2022}, but often the training and inference methods used do not inherently ensure good calibration.
The method presented in this paper is, to the best of our knowledge, the first to consider a complete framework for obtaining calibrated epistemic and aleatoric uncertainty predictions on both energy and forces.

In previous work, we have shown how to extend a graph neural network model for predicting formation energy of molecules to also provide calibrated uncertainty estimates that can be decomposed into aleatoric and epistemic uncertainty~\citep{busk2021}. The method works by combining an ensemble of models with mean-variance outputs and applying post hoc recalibration with isotonic regression on data not used for the training.
In this work, we further extend the approach to include calibrated uncertainty on the force predictions.
Specifically, we extend a neural network potential with a probabilistic predictive distribution on energy and forces, and consider a deep ensemble of models~\citep{deepensembles} to express the aleatoric and epistemic uncertainty about the energy and force predictions.
The uncalibrated predictive distributions are then recalibrated post hoc to fit the error distribution on previously unseen data.
An added benefit of this approach is that ensemble models are generally known to produce more accurate predictions than single models~\citep{schran2020}.
Through computer experiments, we demonstrate that the proposed method results in accurate and calibrated predictions on two publicly available datasets, ANI-1x~\citep{ani1x} and Transitions1x~\citep{schreiner2022} containing out-of-equilibrium and near-transition-state structures, respectively.
The main contribution of the work is a complete framework for training and evaluating neural network potentials with accurate predictions and calibrated aleatoric and epistemic uncertainty on both energies and forces.

The rest of the paper is structured as follows.
The proposed method including the extended graph neural network model and the recalibration procedure is described in Section~\ref{sec:methods}.
The datasets, experiments and results are presented in Section~\ref{sec:results}.
Finally, the main findings and perspectives are discussed in Section~\ref{sec:discussion} and we conclude in Section~\ref{sec:conclusion}.

\section{Methods}
\label{sec:methods}

\subsection{Graph neural network model}

As the base model for our ensemble we use PaiNN~\citep{painn}, an equivariant message passing neural network (MPNN) model designed specifically for predicting properties of molecules and materials.
The model provides a mapping from sets of atomic species and positions $\{(Z_i,\vec{r_i})\}$ to potential energy $E$ and interatomic forces $\{\vec{F_i}\}$.
The potential energy is modelled as a sum over the atomic contributions $E_i$:
\begin{equation}
    E = \sum_i E_i \, ,
\end{equation}
and the forces are computed as the derivative of the potential energy with respect to the atomic positions, ensuring conservation of energy:
\begin{equation}
    \vec{F}_i = -\partial E / \partial\vec{r}_i \, .
\end{equation}

Specifically, the model input is represented as a graph, where there is an edge between a pair of atoms if the mutual distance between the atoms is below a certain cutoff.
The cutoff distance is treated as a hyperparameter and is fixed at 5.0~\AA~in all of our experiments.
The neural network architecture consists of a number of interaction layers, where information, or ``messages'', are exchanged along the edges of the input graph to update the hidden node states, followed by a readout function represented by a fully connected neural network that outputs the atom-wise quantities.
The number of interaction layers and the size of the node hidden states are hyperparameters of the model.

\subsection{Extended model with aleatoric uncertainty}

We extend the base model with additional outputs representing the aleatoric energy uncertainty $\sigma^2_E=\sum_i\sigma^2_{E_i}$ and atom-wise aleatoric force uncertainties $\{\sigma^2_{F_i}\}$.
The atom-wise quantities, $\sigma^2_{E_i}$ and $\sigma^2_{F_i}$, are constrained to be positive by passing them through a softplus activation function, $\log(1+\exp(\cdot))$, and adding a small constant for numerical stability.
Note that here we chose to represent the atom-wise aleatoric force uncertainty by a single scalar even though the force vectors are 3-dimensional.
This simplifying assumption means that we consider the noise scale in the spatial dimensions to be isotropic, i.e., uniform in all directions.
Other options would be to represent the aleatoric force uncertainty by a common scalar for all atoms or as atom-wise vectors representing the uncertainty in each direction. However we found the isotropic approach to be a reasonable compromise and to work well in practice and did not study the other solutions further.

\subsection{Model training procedure}

Each network in the ensemble is initialized with random weight parameters $\theta$ and trained individually on the same training dataset using a loss function composed of a weighted sum of the energy and force loss terms:
\begin{equation}
    \mathcal{L}(\theta) = \lambda_E\mathcal{L}_E(\theta) + \lambda_F\mathcal{L}_F(\theta) \, ,
\end{equation}
where the weight $\lambda_F$ is between 0 and 1 and $\lambda_E = (1 - \lambda_F)$.

ML potentials are usually trained with mean squared error (MSE) loss for both the energy and forces.
Using a negative log likelihood (NLL) loss function provides a natural way of training mean-variance models that also consider uncertainty~\cite{nix}.
The mean squared error (MSE) loss for the energy is straight forward. 
The MSE loss for the forces is evaluated per atom and component-wise over the spatial dimensions and is then averaged over the number of atoms.
The negative log likelihood (NLL) loss for energy, assuming a normally distributed error, is given for a single instance by the following expression where $x=\{(Z_i,\vec{r_i})\}$ represents the model input and the observed values of energy and forces are denoted by $E^{\text{obs}}$ and $F^{\text{obs}}$, respectively:
\begin{align}
\text{NLL}_E(\theta) &=
-\log p(E^{\text{obs}}|x,\theta) \\
&= 
\frac{1}{2} \Bigg( \frac{\big( E^{\text{obs}}-E(x)\big)^2}{\sigma_E^2(x)}
+ \log \sigma_E^2(x) + \log2\pi \Bigg)
\, .
\label{eq:nll_energy}
\end{align}
The instance-wise energy losses are then averaged over the number of instances.

Analogous to the MSE loss for forces, the NLL loss for forces is evaluated per atom $i$ and component-wise over the spatial dimensions $D$ (recall that the predicted atom-wise force uncertainty $\sigma_{F_i}^2$ is a single scalar applied over all spatial dimensions):
\begin{align}
\text{NLL}_{F_i}(\theta) &=
\sum_{d=1}^D -\log p(F^{\text{obs}}_{i,d}|x,\theta) \\
&= 
\sum_{d=1}^D
\frac{1}{2} \Bigg( \frac{\big( F^{\text{obs}}_{i,d}-F_{i,d}(x)\big)^2}{\sigma_{F_i}^2(x)}
+ \log \sigma_{F_i}^2(x) + \log2\pi \Bigg)
\, .
\label{eq:nll_force}
\end{align}
The atom-wise force losses are then averaged over the total number of atoms.
Note that NLL with fixed variance is equivalent to (scaled) MSE and the $\log2\pi$ terms are constant and can be omitted in training.
Here, for models that are trained with a combination of MSE and NLL loss on either energy or forces, we scale the NLL loss by the expected uncertainty (determined empirically) to avoid the NLL loss dominating.

Training directly with NLL loss can be unstable due to interactions between the mean and variance in the loss function, so we apply a training procedure similar to previous work~\citep{busk2021}, where the model is always trained with MSE loss for an initial warmup period before linearly interpolating to the NLL loss.
Other more sophisticated methods for training with NLL loss exist~\citep{skafteReliableTrainingEstimation2019a, seitzer2022betanll}, but we found this simple approach to be sufficient to achieve training stability in our experiments.

\subsection{Ensemble model with epistemic uncertainty}

To estimate the epistemic uncertainty, we follow the approach of~\citet{deepensembles} and make an ensemble approximation by combining the predictions of $M$ individual models.
Using a Bayesian interpretation of deep ensemble models~\cite{wilson2020, hoffmann2021, gustaffson2020}, we can interpret the model weights $\theta^{(m)}$ of each ensemble member $m$ as samples from an approximate posterior distribution $q(\theta) \approx p(\theta | \mathcal{D})$, where $\mathcal{D}$ is the training data.
For a regression model with input $x$ and output $y$ trained on a dataset $\mathcal{D}$ we have:
\begin{align}
	p(y|x, \theta) &= \int p(y|x, \theta)p(\theta| \mathcal{D})d\theta, \\
	&\approx \frac{1}{M} \sum_{m=1}^M p(y|x, \theta^{(m)}),\quad \theta^{(m)} \sim p(\theta | \mathcal{D}),\\
	&\approx \frac{1}{M} \sum_{m=1}^M p(y|x, \theta^{(m)}),\quad \theta^{(m)} \sim q(\theta).
\end{align}
The first approximation is to estimate the integral with $M$ samples from the distribution $p(\theta | \mathcal{D})$ and the second approximation comes from approximating the true posterior $p(\theta | \mathcal{D})$ with the distribution $q(\theta)$. The uncertainty arising from $p(y|x, \theta)$ is the aleatoric uncertainty, while the epistemic uncertainty is modeled as the uncertainty arising from the distribution of the model parameters $q(\theta)$.

When applying this interpretation to a ML potential energy model,
the underlying assumption of the model is that energy and force observations are generated by first sampling $\theta$ and then sampling the normally distributed noise, i.e., for a single molecular energy and atomic force we then have:
\begin{align}
    \theta &\sim q(\theta),\\
    E^\text{obs}(x) &\sim \mathcal{N}\left(E_\theta(x), \sigma_{\theta,E}^2(x)\right),\\
	\vec{F}^{\text{obs}}_{i}(x) &\sim \mathcal{N}\left(- \frac{\partial E_\theta}{\partial \vec{r}_i}(x), \mathbf{I} \sigma^2_{\theta,F_i}(x)\right)\ .
\end{align}
We have two levels of stochastic variables, so to calculate the variance we can use the law of total variance:
\begin{equation}
	\Var\left[Y\right] = \underbrace{\E\left[\Var\left[Y|X\right]\right]}_{\text{aleatoric}} + \underbrace{\Var\left[ \E\left[Y|X\right] \right]}_{\text{epistemic}} \ .
\end{equation}
Using the law of total variance, we get the following expression for the observed energy variance:
\begin{align}
	\Var\left[ E^{\text{obs}} \right] &= \E\left[\Var\left[E^{\text{obs}}|\theta \right]\right] + \Var\left[ \E\left[E^{\text{obs}}|\theta\right] \right]\ ,\\
	&= \underbrace{\E_\theta \left[ \sigma^2_{\theta,E}(x) \right]}_{\text{aleatoric}} + \underbrace{\Var_\theta \left( E_\theta(x) \right)}_{\text{epistemic}}\ .
	\label{eq:energy_var}
\end{align}
Since the force observation is a vector, we compute the total variance element-wise:
\begin{align}
	\Var\left[ F^{\text{obs}}_{i,d}, F^{\text{obs}}_{i,d} \right] &= \E\left[\Var\left[F^{\text{obs}}_{i,d},F^{\text{obs}}_{i,d}|\theta \right]\right] \\
	&\phantom{=}+ \Var\left[ \E\left[F^{\text{obs}}_{i,d}|\theta\right] , \E\left[F^{\text{obs}}_{i,d}|\theta\right]\right]\nonumber\\
	&= \underbrace{\E_\theta \left[ \sigma_{\theta,F_i}^2(x)  \right] }_{\text{aleatoric}} + \underbrace{\Var_\theta \left(  - \frac{\partial E_\theta(x)}{\partial r_{i,d}}, -\frac{\partial E_\theta(x)}{\partial r_{i,d}}\right)}_{\text{epistemic}} \, .
	\label{eq:force_var}
\end{align}

Treating the parameters of the ensemble member as samples from an approximate posterior $q(\theta)$ and using the samples to approximate the expectations, we get the following expressions for the energy mean and variance:
\begin{align}
E^{(*)} &= \frac{1}{M} \sum_{m=1}^M E^{(m)}  \, , \\
\sigma_{E^{(*)}}^2 &=
\underbrace{\frac{1}{M} \sum_{m=1}^M \sigma^2_{E^{(m)}}}_{\text{aleatoric}} + 
\underbrace{\frac{1}{M} \sum_{m=1}^M \left( E^{(m)} - E^{(*)}\right)^2}_{\text{epistemic}}  \, .
\label{eq:ensemble-energy-uncertainty}
\end{align}
Similarly, the mean and variance of the forces for a single atom $i$ are given by the following expressions:
\begin{align}
\vec{F}_i^{(*)} &= 
\frac{1}{M} \sum_{m=1}^M \vec{F}_i^{(m)} \, , \\
\sigma^{2}_{F_i^{(*)}} &=
\underbrace{ \frac{1}{M} \sum_{m=1}^M \sigma^{2}_{F_i^{(m)}} }_{\text{aleatoric}} +
\underbrace{ \frac{1}{M} \sum_{m=1}^M \frac{1}{D} \left\| \vec{F}_i^{(m)} - \vec{F}_i^{(*)} \right\|^2 }_{\text{epistemic}} \, ,
\label{eq:ensemble-force-uncertainty}
\end{align}
where $D$ denotes the spatial dimensions. With the assumption of isotropic force variance, we average across the spatial dimension in \eqref{eq:ensemble-force-uncertainty} when estimating the epistemic force variance of \eqref{eq:force_var} using the sample variance.
The mean energy and forces represent the ensemble prediction and the energy and force variances represent the ensemble total uncertainties, which can be decomposed into aleatoric and epistemic components as shown above.

When we want to evaluate the likelihood of an observation, we need access to the predictive distribution and knowing the mean and variance is not sufficient. Following~\citet{deepensembles}, we parameterize a normal distribution with the mean and variance. With the mean and variance specified, the normal distribution is the maximum entropy probability distribution, i.e., we make the least assumptions about the data by using a normal distribution following the maximum entropy principle~\citep{jaynes1957,blowerInformationProcessingMaximum2013}. However, to hold true, this would require all the variance outputs of the ensemble members to be equal. If the variances follow an inverse-gamma distribution, the predictive distribution would be a student-t in the infinite ensemble limit, which is the assumption used in deep evidential regression~\citep{amini2020, skafteReliableTrainingEstimation2019a}.

\subsection{Uncertainty calibration}

Several methods exist for evaluating the calibration of regression models.
NLL provides a standard metric for quantifying the overall quality of probabilistic models by measuring the probability of observing the data given the predicted distribution.
However, NLL depends on both the predicted mean and uncertainty (see eq.~\ref{eq:nll_energy} and \ref{eq:nll_force}) and it can be useful to evaluate only the quality of the uncertainty estimates.
For example, it is often informative to visually compare the predicted uncertainties with the empirical errors by plotting them.
Since the total uncertainty of the ensemble model (eq.~\ref{eq:ensemble-energy-uncertainty} and eq.~\ref{eq:ensemble-force-uncertainty}) can be interpreted as a variance of a normal distribution, we expect most of the errors to lie within 2-3 standard deviations of the predictive distribution.
The variance-based approach also allows us to evaluate standard scores, also known as z-scores, defined as the empirical error divided by the standard deviation of the predictive distribution~\citep{pernot2022}:
\begin{equation}
    z = \frac{y^{\text{obs}}-y(x)}{\sigma(x)} \, .
\end{equation}
A z-score variance (ZV) close to 1 indicates that the predicted uncertainty on average corresponds to the variance of the error and thereby is an indication of good average calibration.
The same approach can be applied to subsets of the data leading to a local z-score variance (LZV) analysis, for example by evaluating ZV in equal size bins of increasing uncertainty to test the consistency of the uncertainty estimates~\citep{pernot2022}.
For plotting, we found it useful to report the square root of the z-variance (RZV).
Additionally, we can assess how well the uncertainty estimates correspond to the expected error locally by sorting the predictions in equal size bins by increasing uncertainty and plotting the root mean variance (RMV) of the uncertainty versus the empirical root mean squared error (RMSE), also known as an error-calibration plot or reliability diagram.
The error-calibration can be summarized by the expected normalized calibration error (ENCE)~\citep{levi2022}, which measures the mean difference between RMV and RMSE normalised by RMV:
\begin{equation}
    \text{ENCE} = \frac{1}{K}\sum_{k=1}^K \frac{|\text{RMV}_k-\text{RMSE}_k|}{\text{RMV}_k} \, ,
\end{equation}
where $k=1,\dots,K$ iterates the bins.
LZV analysis and the reliability diagram provide two useful ways to evaluate the local consistency of the uncertainty estimates.
We use 15 equal sized bins in all of our analyses.

Good average or local calibration is not sufficient to ensure that individual uncertainty estimates are informative, i.e.,~if the uncertainty estimates are homoscedastic, they are not very useful.
Thus it is generally desirable for uncertainty estimates to be as small as possible while also having some variation, which is also known as sharpness.
To measure sharpness, we report the root mean variance (RMV) of the uncertainty, which should be small and correspond to the RMSE, and the coefficient of variation (CV), which quantifies the ratio of the standard deviation of the uncertainties with the mean uncertainty and thus the overall dispersion, or heteroscedasticity, of the predicted uncertainty:
\begin{equation}
    \text{CV} = \frac{\sqrt{N^{-1} \sum_{n=1}^N (\sigma(x_n) - \overline\sigma)^2}}{\overline\sigma} \, ,
\end{equation}
where $n=1,\dots,N$ in this case iterates the test dataset, $\sigma(x_n)$ is the predicted standard deviation (uncertainty) of instance $n$ and
$\overline\sigma = N^{-1} \sum_{n=1}^N \sigma(x_n)$
is the mean predicted standard deviation.
If uncertainties are heteroscedastic while having good local calibration, it is also an indication of good ranking ability which is important in certain applications such as active learning~\citep{schran2020}.

\subsection{Uncertainty recalibration}
\label{sec:recalibration}

The model training procedure described above does not by itself ensure good uncertainty calibration when the model is applied to unseen data.
The individual models may overfit to the training data and the total ensemble uncertainties (eq. \ref{eq:ensemble-energy-uncertainty} and \ref{eq:ensemble-force-uncertainty}) are strictly greater than any of the individual model uncertainties, and not fitted on any data.
Therefore, following the approach of our previous work, \citet{busk2021}, we recalibrate the ensemble uncertainty estimates post hoc by using a recalibration function that maps the uncalibrated predictive distribution to a calibrated distribution.
The recalibration function is a non-linear uncertainty scaling function based on isotonic regression fitted to predict empirical squared errors on the validation set.
Specifically, the recalibration function maps the uncalibrated uncertainty estimates $\sigma^2$ to scaled uncertainty estimates $s^2\sigma^2$, where $s^2$ is the predicted scaling factor.
In our experiments, both energy uncertainties $\sigma^2_{E^{(*)}}$ and force uncertainties $\sigma^2_{F^{(*)}}$ are recalibrated in this way.

Because the recalibration function is a scaling function, the recalibration procedure does not change the mean of the predictive distribution and thus does not change the prediction.
Additionally, the isotonic regression results in a monotonic increasing scaling function and thus preserves the ordering of the uncertainty estimates and thereby the ranking.
We use the implementation of isotonic regression available in the scikit-learn Python package~\cite{scikit-learn}.

\section{Experiments and Results}
\label{sec:results}

\begin{table}
\footnotesize
\caption{Test results after recalibration of ensemble models ($M=5$) trained on the ANI-1x (A1x) and Transition1x (T1x) datasets with different combinations of mean squared error (MSE) and negative log likelihood (NLL) loss functions on the energies and forces.
Energy errors are averaged over molecules, while force errors are computed component-wise and averaged over the spatial dimensions and atoms.}
\label{tab:results}
\centering
\makebox[\textwidth]{\begin{tabular}{@{\extracolsep{\fill}}rllrrrrrrrrrrrrrr}
\toprule
\multicolumn{1}{c}{\bf Data} & \multicolumn{2}{c}{\bf Loss} & \multicolumn{7}{c}{\bf Energy (eV)} & \multicolumn{7}{c}{\bf Forces (eV/\AA)} \\
\cmidrule(r){2-3}\cmidrule(lr){4-10}\cmidrule(l){11-17}
& $\mathcal{L}_E$ & $\mathcal{L}_F$ & MAE$\downarrow$ & RMSE$\downarrow$ & NLL$\downarrow$ & RZV & ENCE$\downarrow$ & CV & RMV & MAE$\downarrow$ & RMSE$\downarrow$ & NLL$\downarrow$ & RZV & ENCE$\downarrow$ & CV & RMV \\
\midrule
\multirow{4}*{A1x} & MSE & MSE &
0.0123 & 0.0278 & -2.81 & 0.97 & 0.0773 & 1.27 & 0.0283 &
0.0180 & 0.0362 & -2.56 & 0.98 & 0.0243 & 1.24 & 0.0386 \\
& NLL & MSE &
0.0118 & 0.0256 & -3.04 & 1.00 & 0.0411 & 1.00 & 0.0209 &
0.0179 & 0.0364 & -2.56 & 0.98 & 0.0229 & 1.22 & 0.0381 \\
& MSE & NLL &
0.0117 & 0.0305 & -2.91 & 0.97 & 0.0928 & 1.53 & 0.0305 &
0.0175 & 0.0399 & -2.77 & 1.00 & 0.0099 & 1.51 & 0.0410 \\
& NLL & NLL &
0.0105 & 0.0296 & \bf{-3.26} & 1.02 & 0.0600 & 1.51 & 0.0237 &
0.0171 & 0.0402 & \bf{-2.79} & 1.00 & 0.0093 & 1.56 & 0.0409 \\
\midrule
\multirow{4}*{T1x} & MSE & MSE &
0.0344 & 0.0612 & -1.79 & 0.89 & 0.1144 & 0.82 & 0.0682 &
0.0370 & 0.0743 & -1.94 & 0.99 & 0.0292 & 1.21 & 0.0804 \\
& NLL & MSE &
0.0318 & 0.0578 & -1.94 & 0.92 & 0.1383 & 0.86 & 0.0655 &
0.0366 & 0.0744 & -1.97 & 0.98 & 0.0293 & 1.27 & 0.0831 \\
& MSE & NLL &
0.0332 & 0.0600 & -1.84 & 0.95 & 0.1108 & 0.83 & 0.0628 &
0.0369 & 0.0745 & -1.99 & 0.94 & 0.0615 & 1.18 & 0.0817 \\
& NLL & NLL &
0.0303 & 0.0574 & \bf{-2.09} & 0.98 & 0.0906 & 0.96 & 0.0562 &
0.0359 & 0.0751 & \bf{-2.05} & 0.94 & 0.0645 & 1.15 & 0.0773 \\
\bottomrule
\end{tabular}}
\end{table}

\subsection{Datasets}

The proposed method was evaluated on two publicly available datasets designed specifically for the development and evaluation of ML potentials, ANI-1x~\citep{ani1x} and Transitions1x~\citep{schreiner2022}.
The datasets include out-of-equilibrium and near-transition-state structures, respectively, and represent varied energy and force distributions.
The ANI-1x dataset consists of Density Functional Theory (DFT) calculations for approximately 5 million diverse molecular conformations with an average of 8 heavy atoms (C, N, O) and an average of 15 total atoms (including H) along with multiple properties including total energy and interatomic forces computed at the \textomega B97x/6-31G(d) level of theory.
The dataset was generated by perturbing equilibrium configurations using an active learning procedure to ensure conformational diversity with the aim of developing an accurate and transferable ML potential.
The Transition1x dataset contains DFT calculations of energy and forces, for 9.6 million molecular conformations with up to 7 heavy atoms (C, N, O) and an average of 14 total atoms (including H), likewise computed at the \textomega B97x/6-31G(d) level of theory. 
Here, the structures were sampled on and around full reaction pathways, thus including conformations far from equilibrium and near transition states.
The dataset was generated by running a nudged elastic band (NEB)~\citep{sheppard2008} algorithm with DFT on a set of approximately 10 thousand organic reactions with up to 6 bond changes while saving intermediate calculations with the aim of improving ML potentials around transition states.
Transition1x is more varied in terms of interatomic distances between pairs of heavy atoms than ANI-1x, but less varied in terms of the distribution of forces, since forces are generally minimised along reaction pathways~\citep{schreiner2022, sheppard2008}.

\subsection{Model hyperparameters}

The same general model configuration was used in all experiments.
Each individual graph neural network model was configured with 3 interaction layers, a hidden node state size of 256 and a 2-layer atom-wise readout network with 3 outputs representing $E_i$, $\sigma^2_{E_i}$ and $\sigma^2_{F_i}$.
The input molecular graphs were generated with an edge cutoff radius of 5.0~\AA.
Models were trained using the Adam optimizer with an initial learning rate of $10^{-4}$, an exponential decay learning rate scheduler, a batch size of 64 molecular graphs, force loss weight $\lambda_F = 0.5$ and an early stopping criterion on the validation loss to prevent overfitting.
In each experiment, models were trained individually with the same hyperparameters on the same training data, but with different random parameter initialisation and random shuffling of the training data to induce model diversity in the ensembles.
For each ensemble model, after the training was completed, a recalibration function was fitted using predictions on the respective validation dataset following the procedure described in Section~\ref{sec:recalibration} and applied to the predictions on the test data.
For ensemble models trained with MSE loss, only the epistemic uncertainty is considered.

\subsection{ANI-1x results}

\begin{figure}[h]
\centering 
\includegraphics[width=0.5\linewidth]{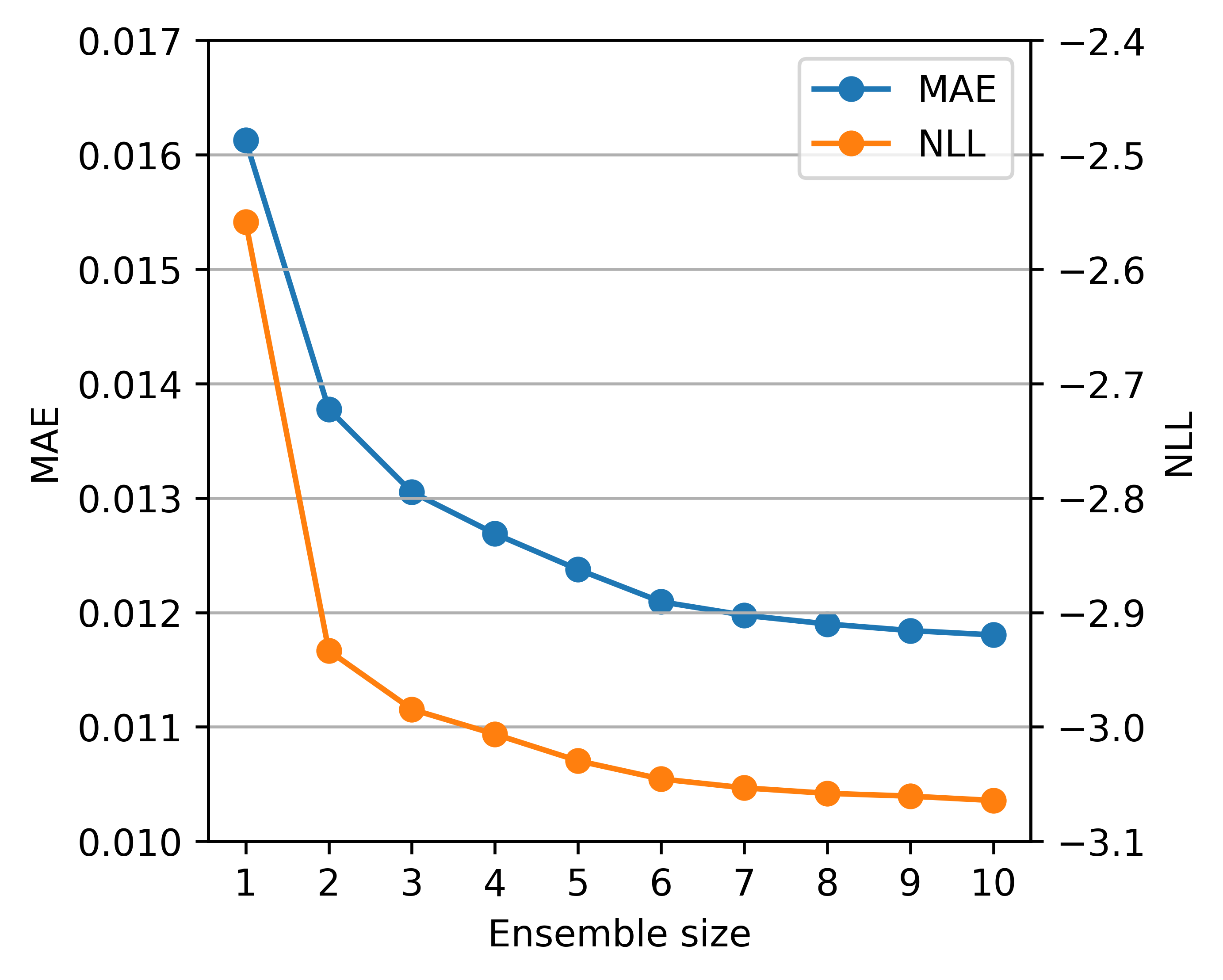}
\caption{Trade-off between performance and ensemble size on the ANI-1x validation dataset using ensembles of models trained with NLL loss on both energy and forces.}
\label{fig:ani1x-ensemble-size}
\end{figure}

Ensembles of graph neural network models were trained on ANI-1x using the data splits from~\citet{schreiner2022}, where the validation and test datasets consist of approximately 5\% of the data each and the training set consists of the remaining 90\% of the data.
The splits are stratified by chemical formula to ensure different splits do not contain configurations made up of exactly the same atoms and selected such that all splits include all species of heavy atoms.
Individual models were trained for up to 10 million gradient steps (approximately 144 epochs) with an initial warmup period of 2 million steps where the model was trained only with MSE loss followed by an interpolation period of 1 million steps, where the loss was interpolated linearly to NLL loss (only NLL models).
The ensemble predictions were then recalibrated post hoc using a recalibration function fitted using the validation dataset.
The trade-off between validation performance and ensemble size $M$ using models trained with NLL loss on both energy and forces is illustrated in Figure~\ref{fig:ani1x-ensemble-size}.
Using a larger ensemble size results in lower error, as expected, but comes at the cost of additional computations.
We observe that a reasonably low error is obtained at $M=5$ and only small improvements are gained beyond that, which is similar to what we found in previous work~\cite{busk2021}.

Test results for $M=5$ ensembles trained with different combinations of MSE and NLL loss on energy and forces are presented in the first four rows of Table~\ref{tab:results}.
The ensemble trained with MSE loss on both energy and forces is a standard ensemble model with post hoc recalibration.
The other three ensembles show the effect of training with NLL loss on either energy or forces or both.
All four ensembles achieved a low error on energy and forces in terms of MAE and RMSE compared to the results reported by~\citet{schreiner2022} using a PaiNN model similar to the base model of our ensembles (MAE=0.023 on energy and MAE=0.039 on forces).
Importantly, training with NLL loss on either energy or forces or both did not result in worse performance in terms of prediction error.

\begin{figure*}[t]
\centering
\includegraphics[height=0.29\linewidth]{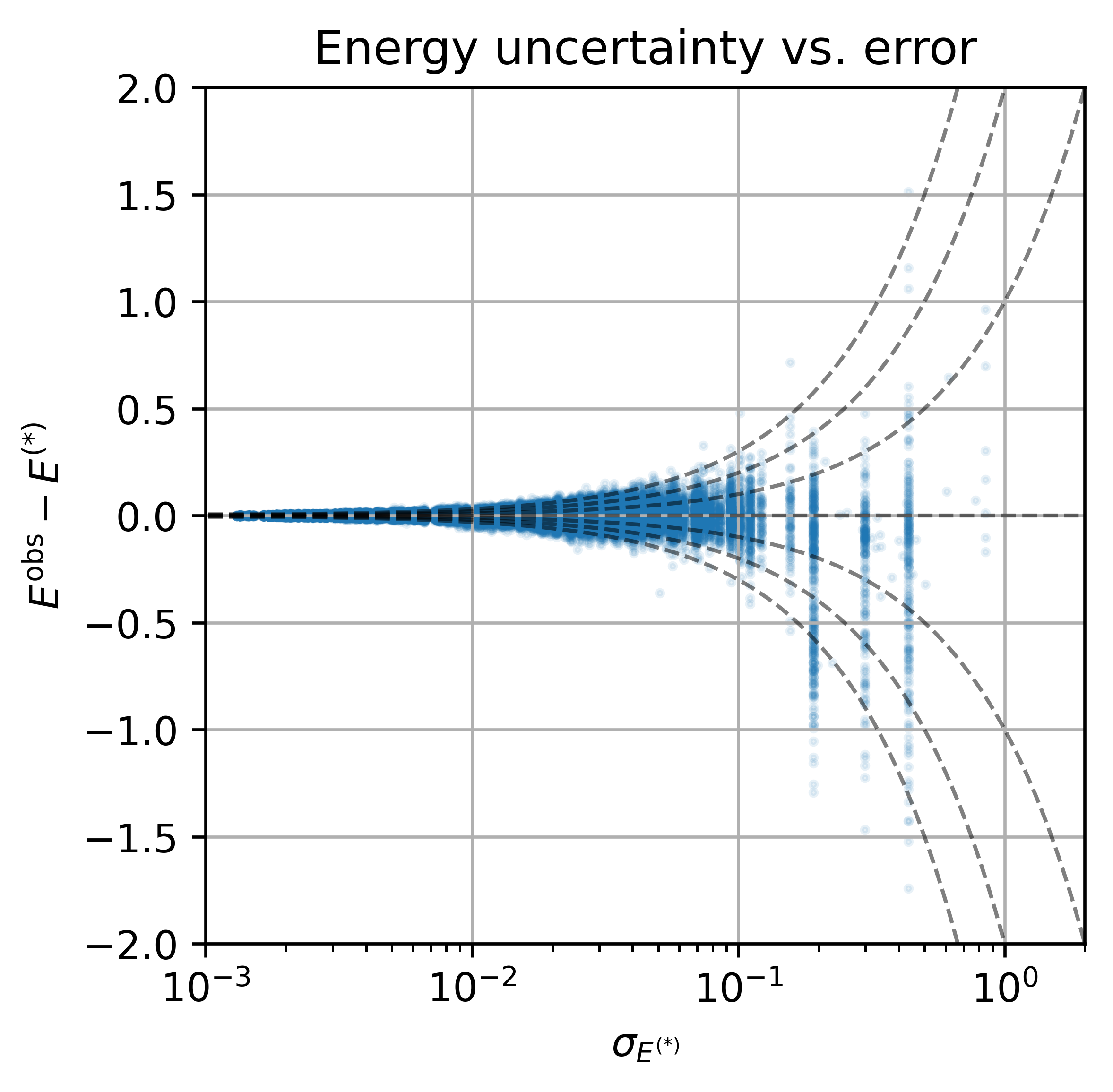}
\includegraphics[height=0.29\linewidth]{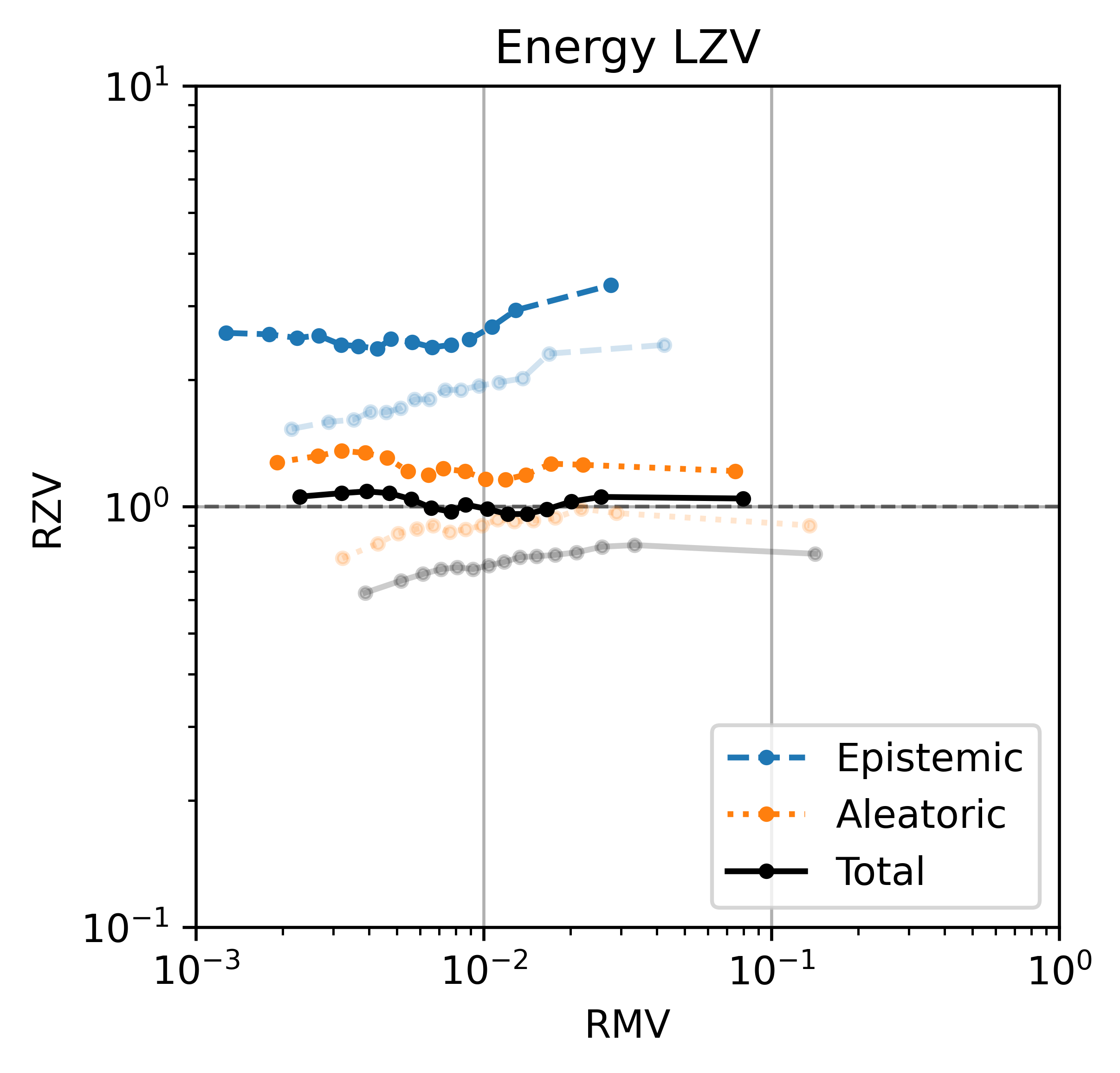}
\includegraphics[height=0.29\linewidth]{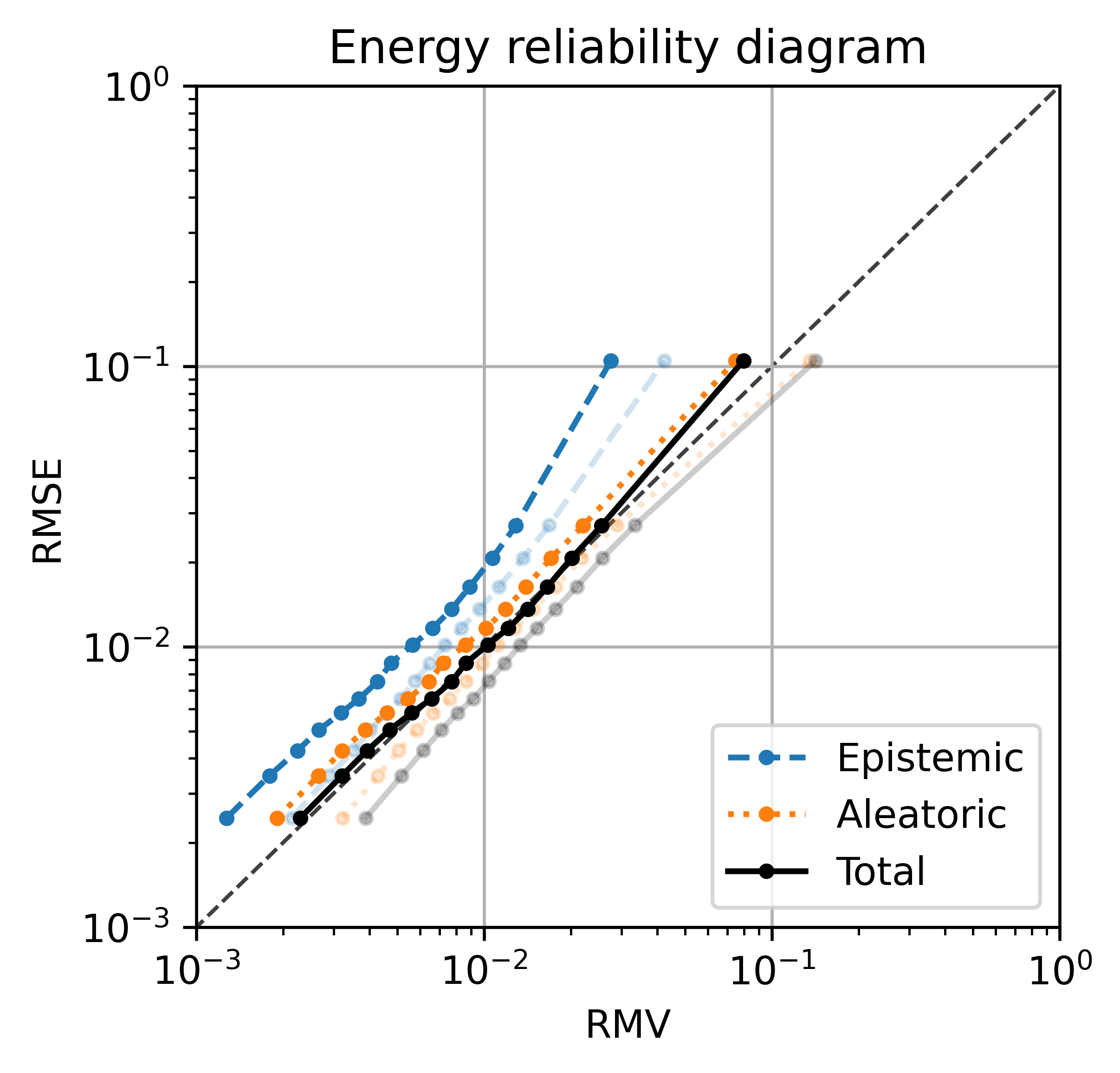}
\\
\includegraphics[height=0.29\linewidth]{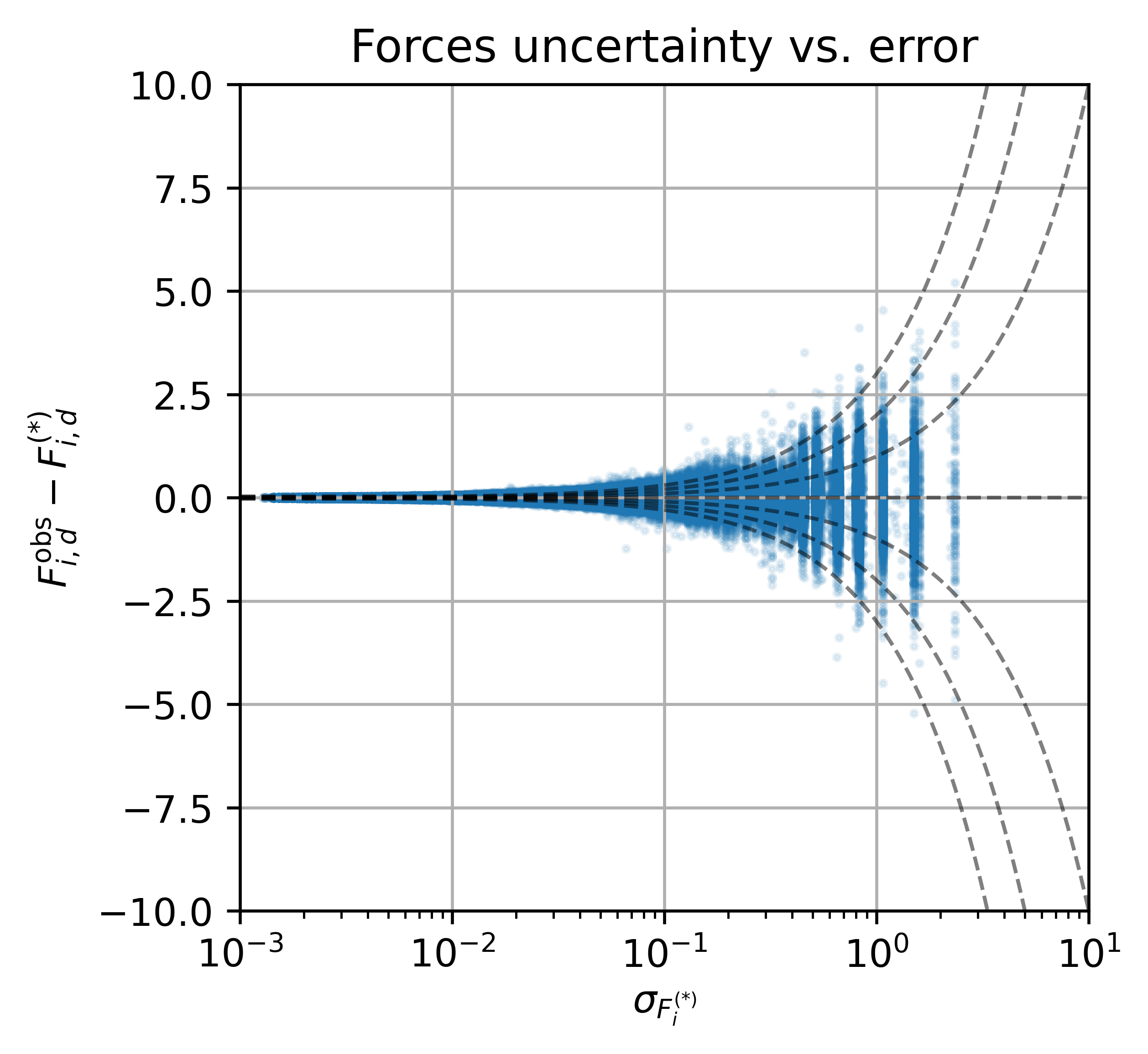}
\includegraphics[height=0.29\linewidth]{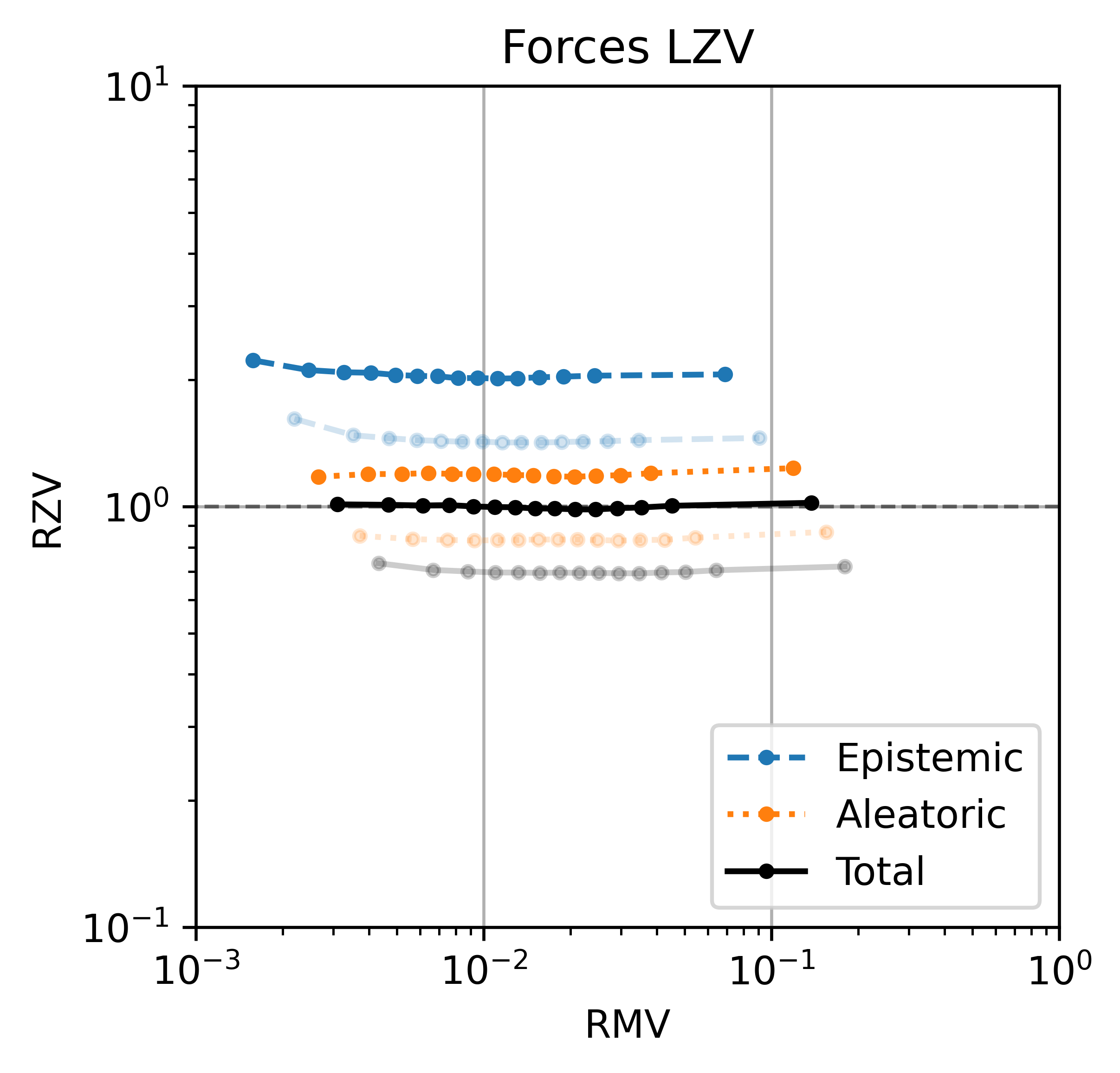}
\includegraphics[height=0.29\linewidth]{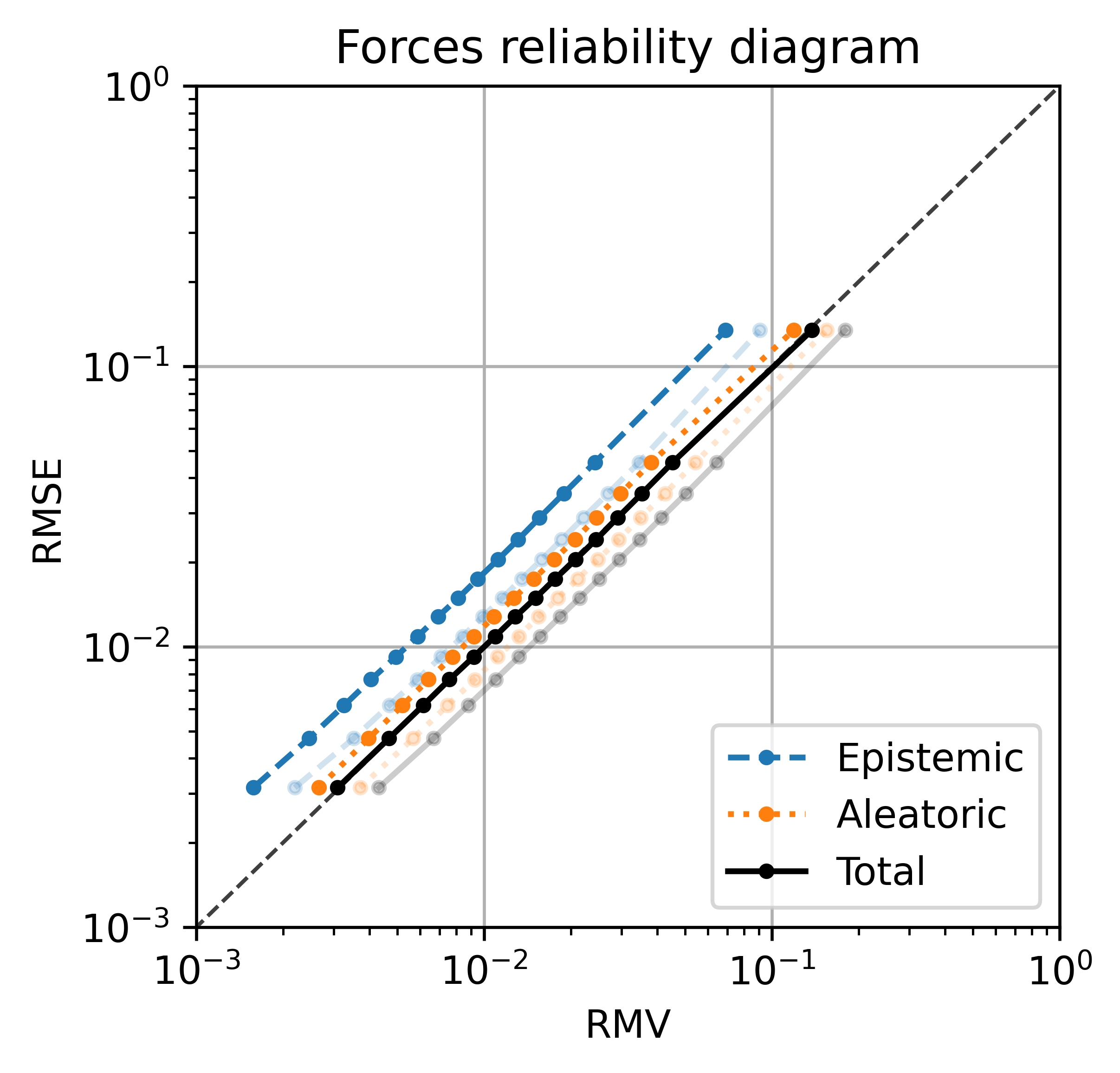}
\caption{Calibration results on the ANI-1x dataset of energy (top row) and forces (bottom row) for an ensemble of $M=5$ models trained with NLL loss on both energy and forces.
To illustrate the effect of recalibration, the transparent curves show results before applying recalibration (energy ENCE=0.2650, forces ENCE=0.2964) whereas the solid curves show results after recalibration (energy ENCE=0.0600, forces ENCE=0.0093).
The LZV analyses and reliability diagrams are generated using 15 equal sized bins.
All curves in each plot use the same bins based on sorting by total uncertainty.
}
\label{fig:calibration-ani1x-nll-nll}
\end{figure*}

All four ensemble models achieved good average calibration on ANI-1x in terms of NLL and RZV after recalibration, but ensembles trained with NLL loss performed slightly better which is observed both on energy and forces and the ensemble trained with NLL on both energy and forces performs best overall.
Additionally, all ensembles scored a high CV indicating uncertainty estimates are heteroscedastic and thus informative.
Calibration plots for the ensemble trained on ANI-1x with NLL loss on energy and forces are presented in Figure~\ref{fig:calibration-ani1x-nll-nll}.
The uncertainty vs.~error plots show that in general large errors are associated with large uncertainties and most errors are within 2-3 standard deviations of uncertainty as desired. 
For the energy, the model appears to be biased for some examples with large errors, but these are relatively few and are correctly identified as problematic by high uncertainty.
For the forces, the distribution of errors looks more symmetrical around zero.
This is also clearly shown by the local z-score analysis plots where for the energy, the variance of the z-scores is slightly off for very low and very high uncertainties, although still centered around 1, whereas for the forces the variance of the z-scores is close to 1 for all uncertainties which indicates high consistency. 
Finally, the reliability diagrams show the relation between predicted uncertainty and expected error for the energy and forces, respectively.
Both plots show a clear correlation between the uncertainty and the expected error as the curves lie close to the identity line.
Again, the model very slightly underestimates the expected error of the energy at low and high uncertainties, and the curve for the forces is near perfect.
The reliability diagrams are summarised by ENCE scores in Table~\ref{tab:results}.
Additional calibration results are included in the ESI.

\subsection{Transition1x results}
\label{sec:t1x-results}

Analogous to the first experiment, ensembles of graph neural network models were trained on Transition1x using data splits from~\citet{schreiner2022} based on the same splitting criteria as ANI-1x described above.
Models were trained for up to 3 million gradient steps (approximately 21 epochs) with an initial warmup period of 2 million steps followed by an interpolation period of 2 million steps (training was stopped before finishing the full interpolation period).
When training for longer on this dataset, we observed severe overfitting. We believe this is because the data was generated from a relatively small set of chemical reactions making the models prone to overfit the many similar configurations associated with the same reactions in the training data.
The ensemble predictions were recalibrated post hoc using a recalibration function fitted using the validation dataset.
Varying the ensemble size yielded similar results to the first experiment (Figure~\ref{fig:ani1x-ensemble-size}) and $M=5$ was selected as a good compromise between performance and computational cost.

\begin{figure*}[t]
\centering
\includegraphics[height=0.29\linewidth]{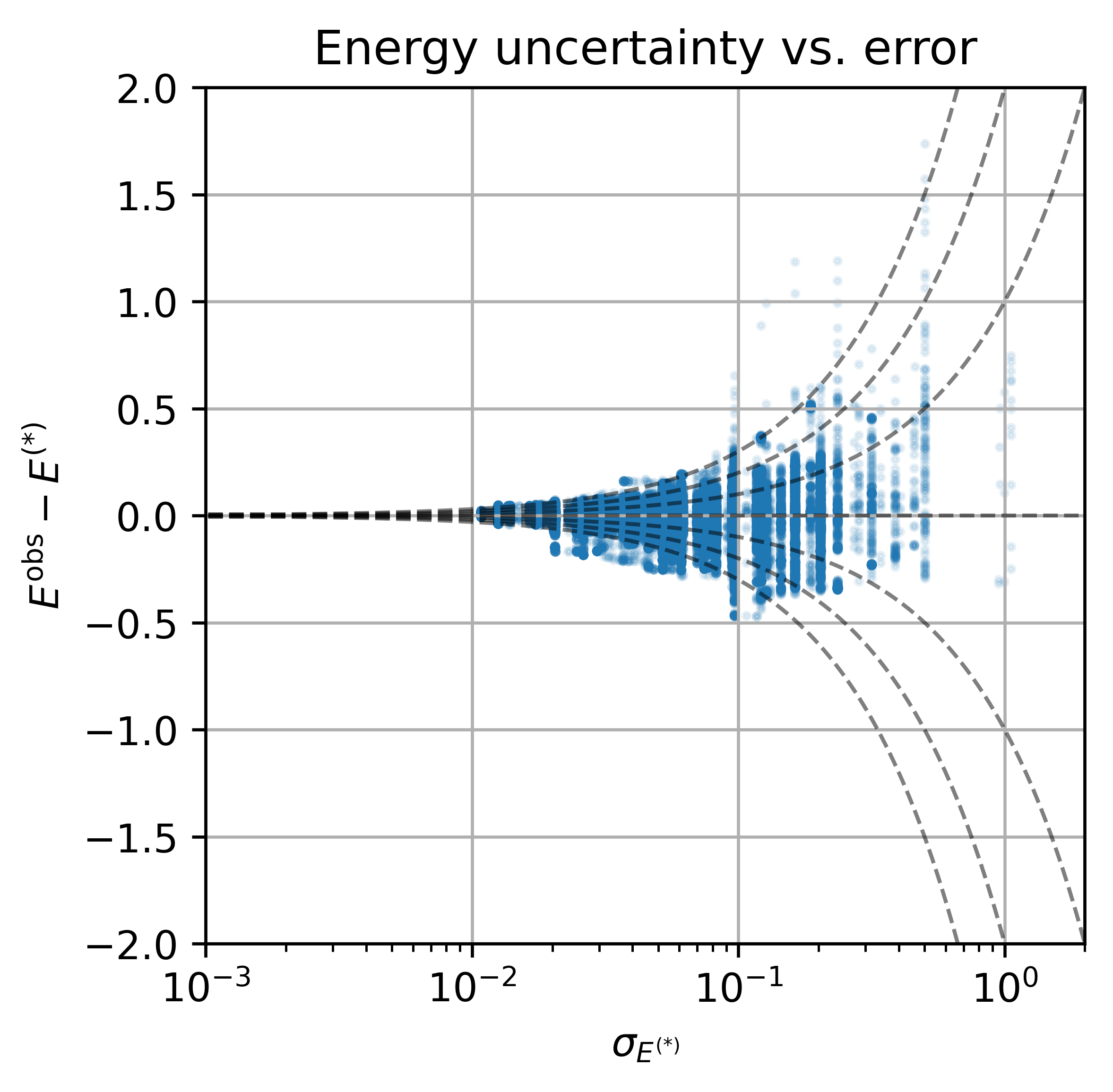}
\includegraphics[height=0.29\linewidth]{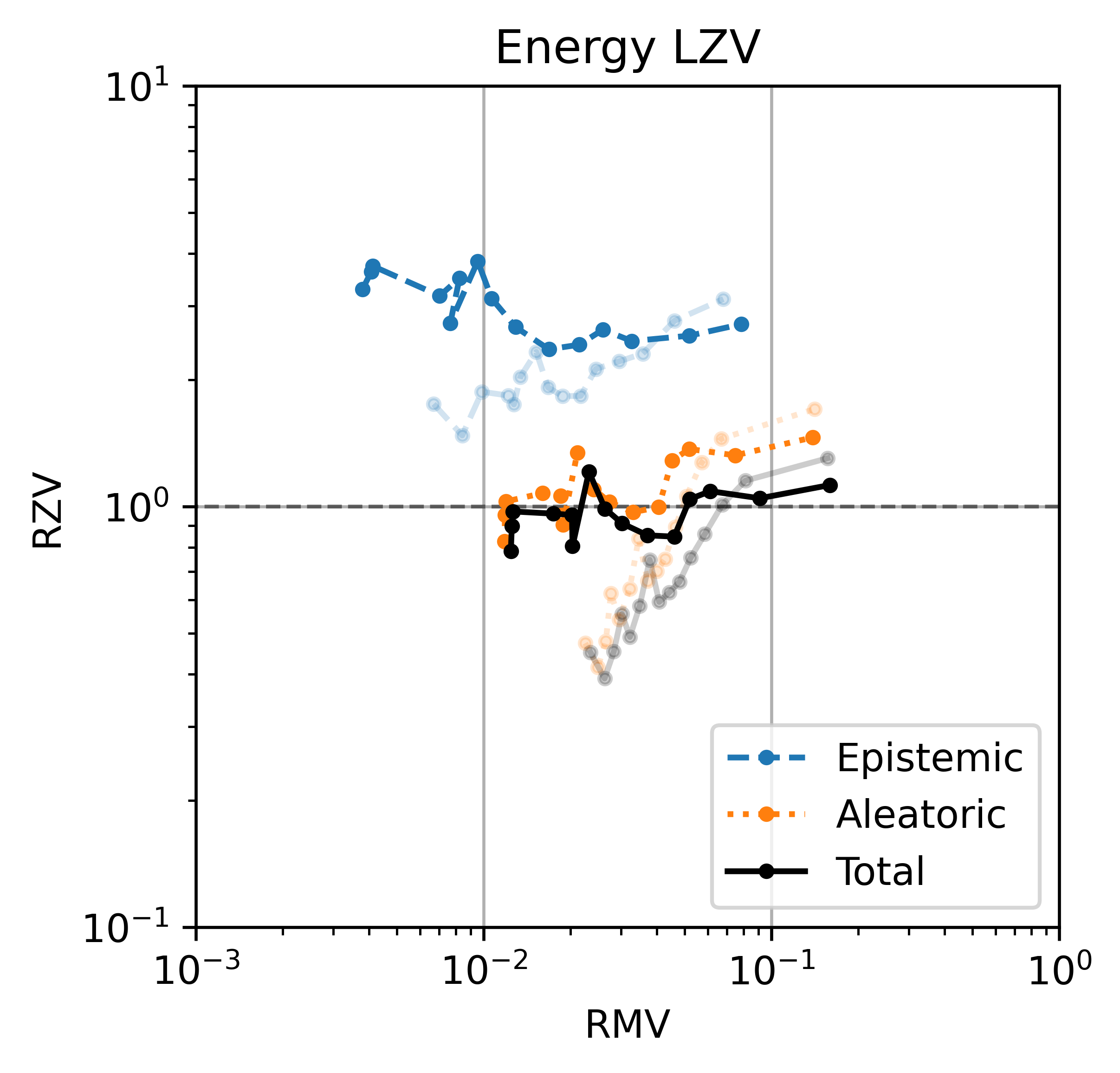}
\includegraphics[height=0.29\linewidth]{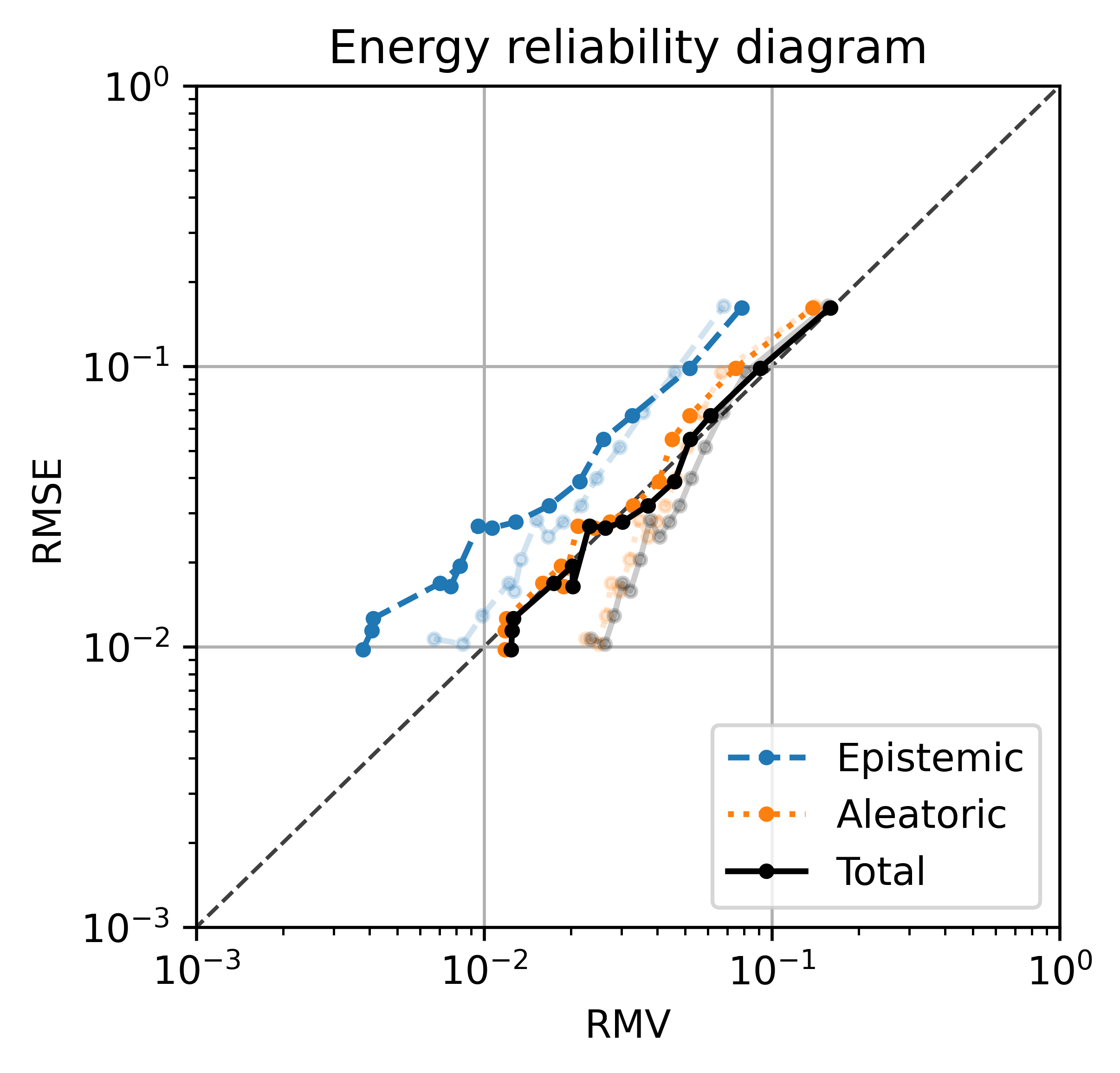}
\\
\includegraphics[height=0.29\linewidth]{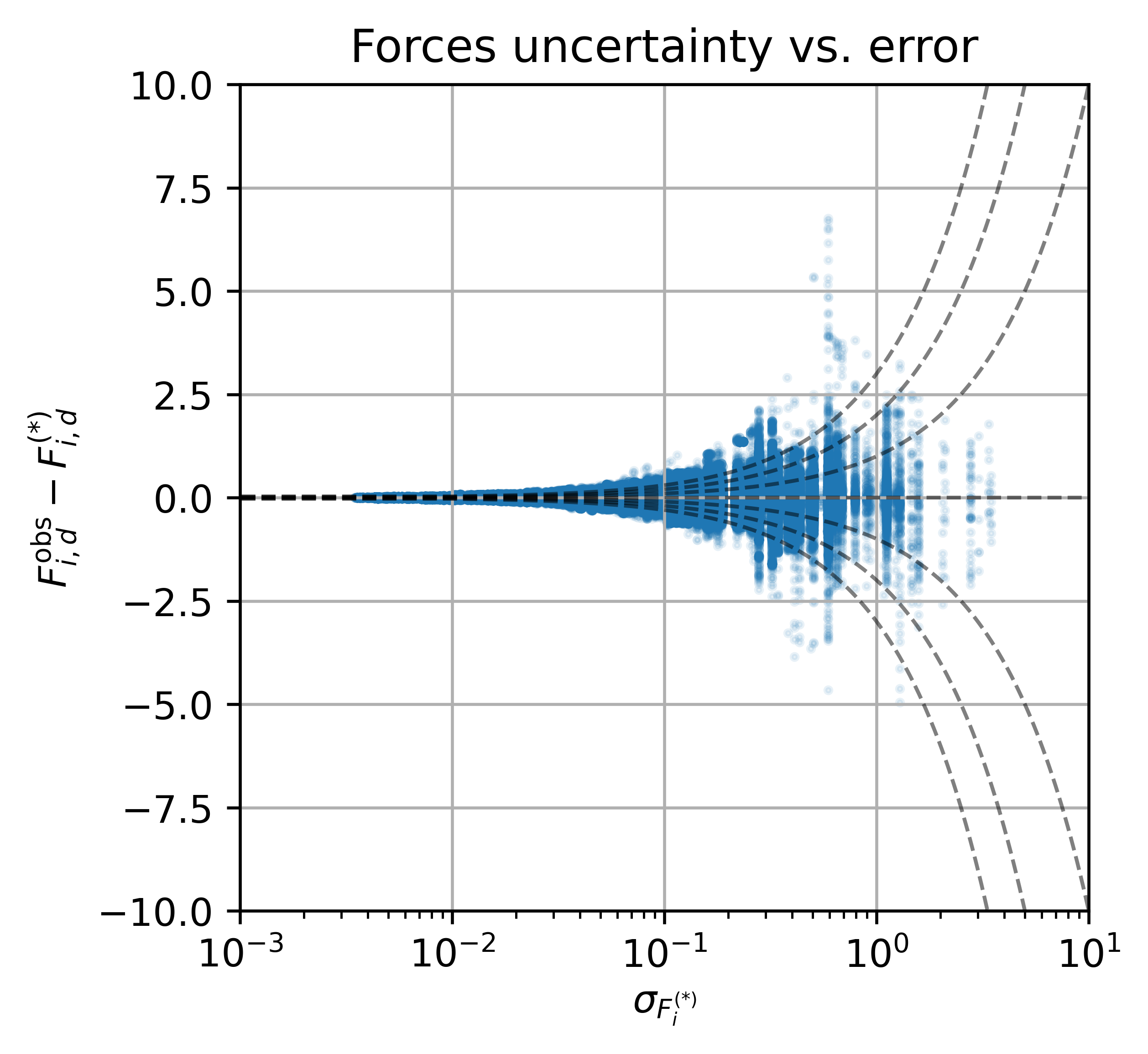}
\includegraphics[height=0.29\linewidth]{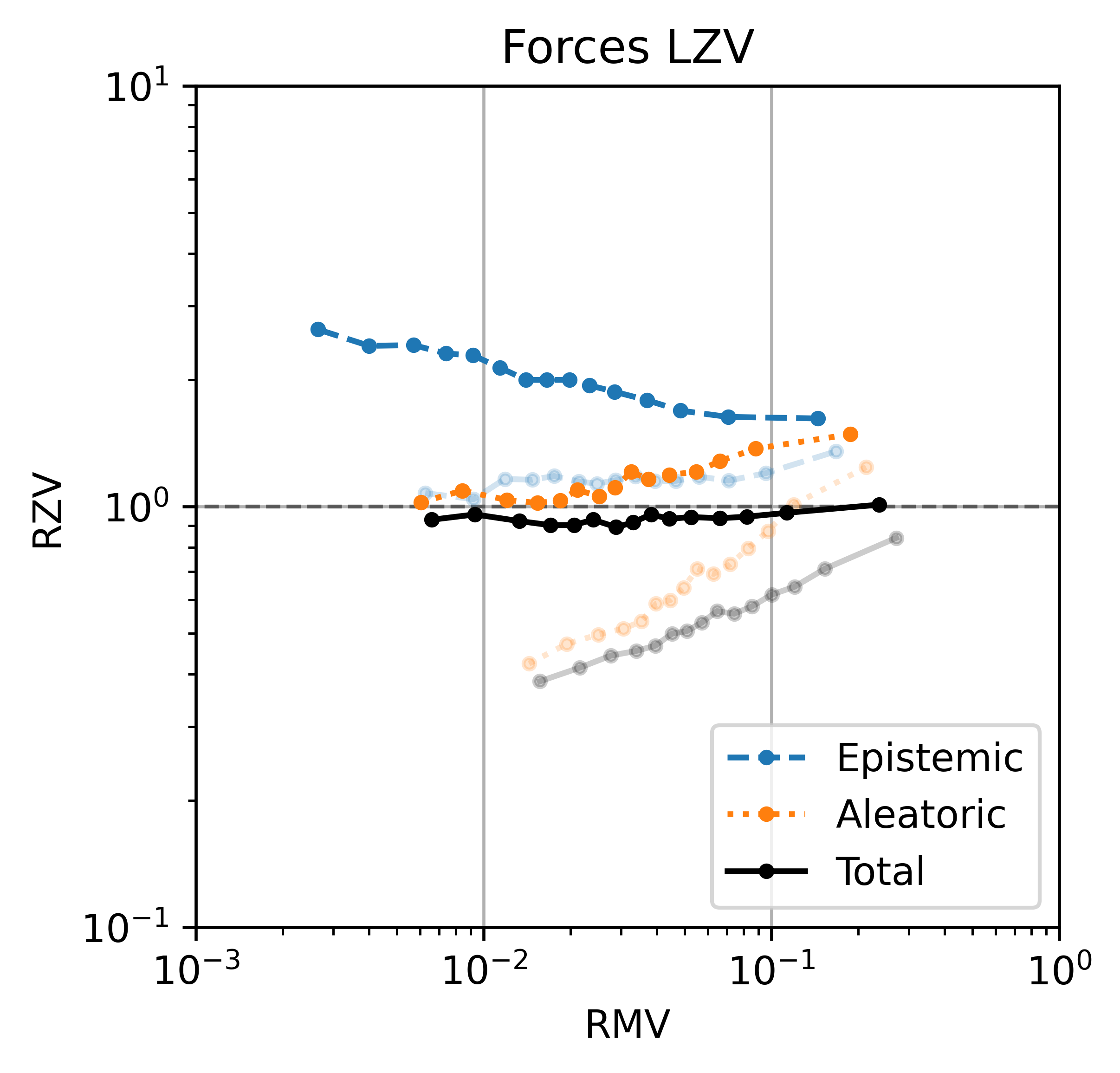}
\includegraphics[height=0.29\linewidth]{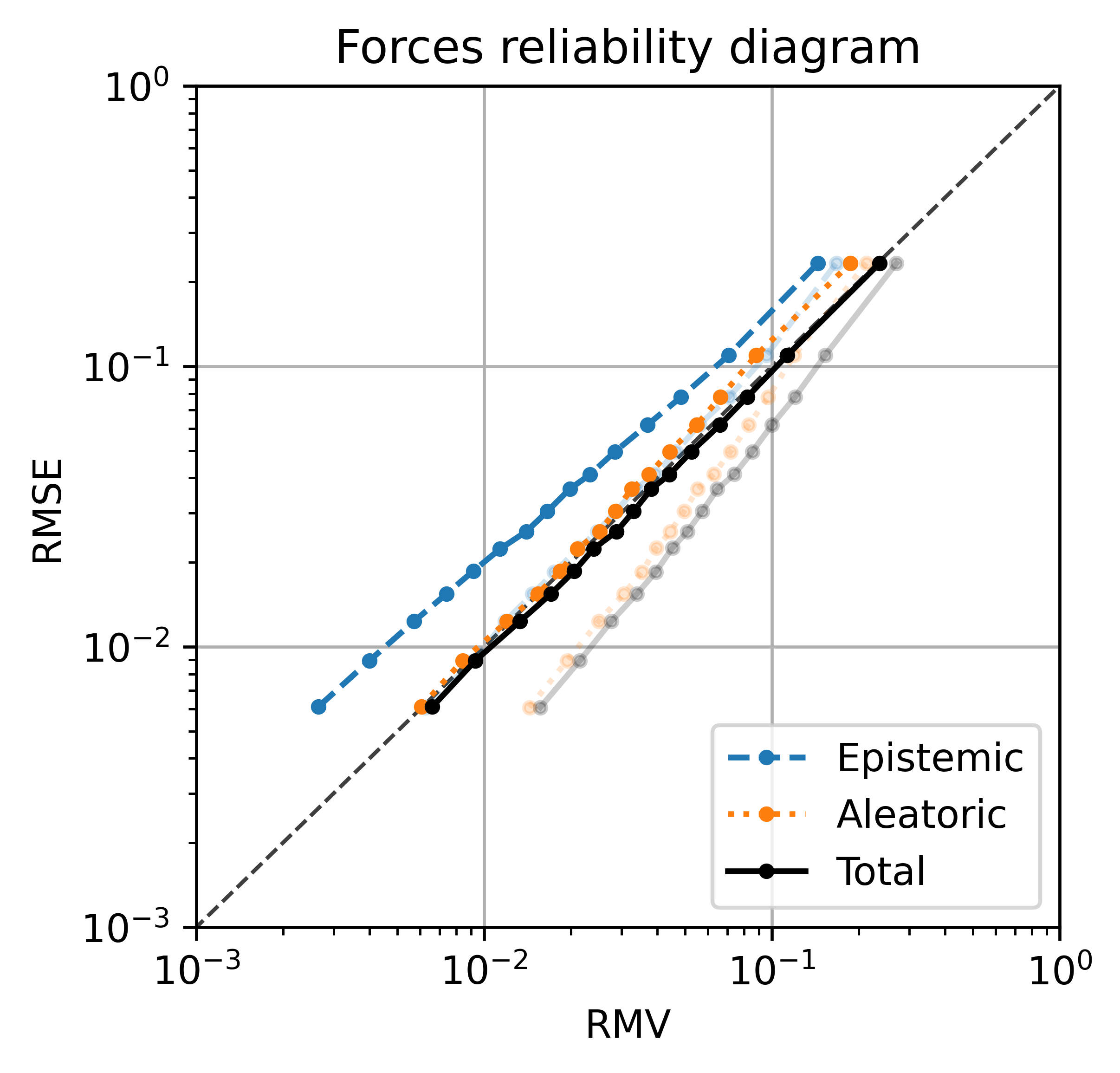}
\caption{Calibration results on the Transition1x dataset of energy (top row) and forces (bottom row) for an ensemble of $M=5$ models trained with NLL loss on both energy and forces.
To illustrate the effect of recalibration, the transparent curves show results before applying recalibration (energy ENCE=0.3339, forces ENCE=0.4502) whereas the solid curves show results after recalibration (energy ENCE=0.0906, forces ENCE=0.0645).
The LZV analyses and reliability diagrams are generated using 15 equal sized bins.
All curves in each plot use the same bins based on sorting by total uncertainty.
}
\label{fig:calibration-t1x-nll-nll}
\end{figure*}

\begin{figure}[h]
\centering 
\includegraphics[width=0.4\linewidth]{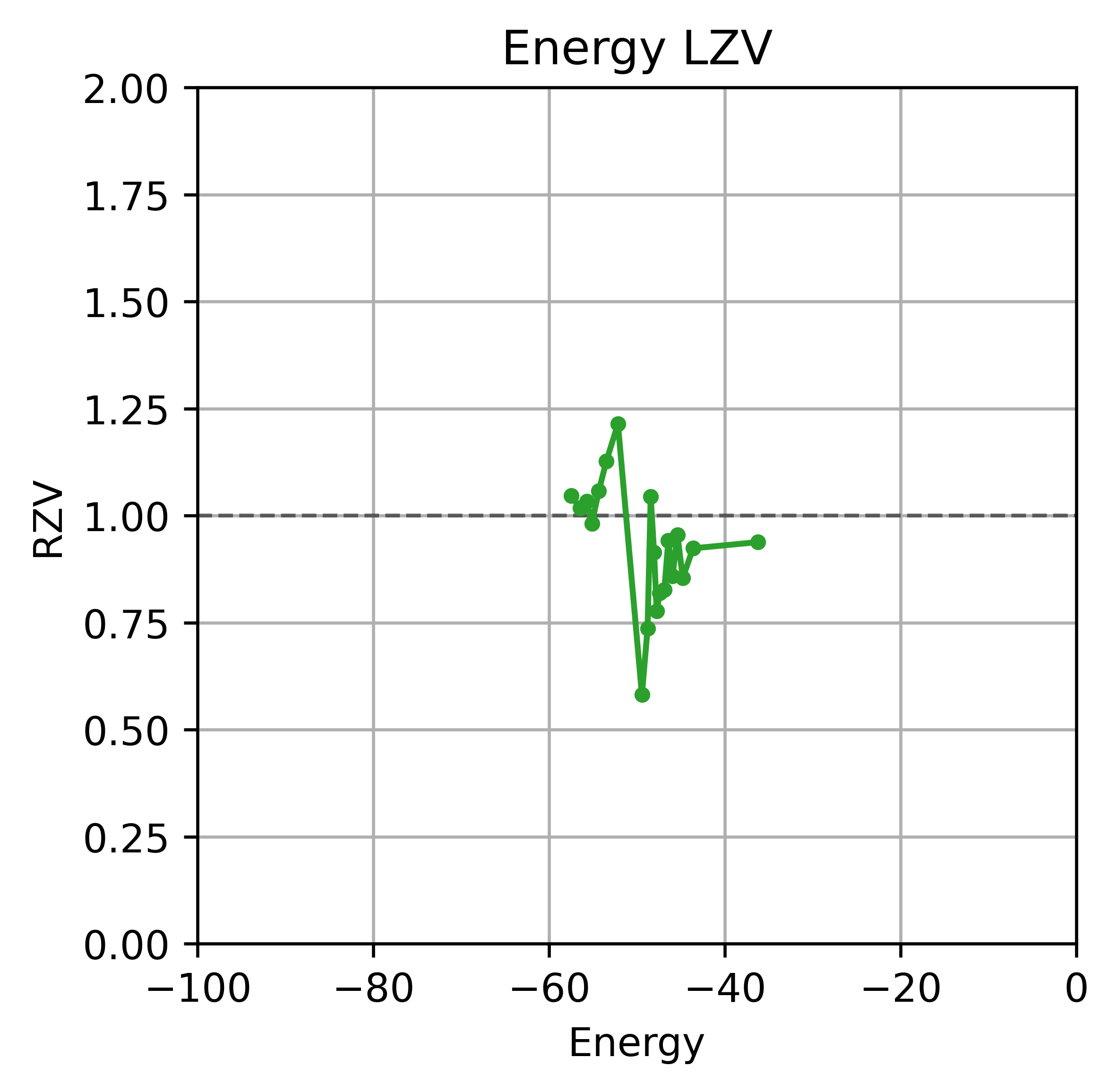} \includegraphics[width=0.4\linewidth]{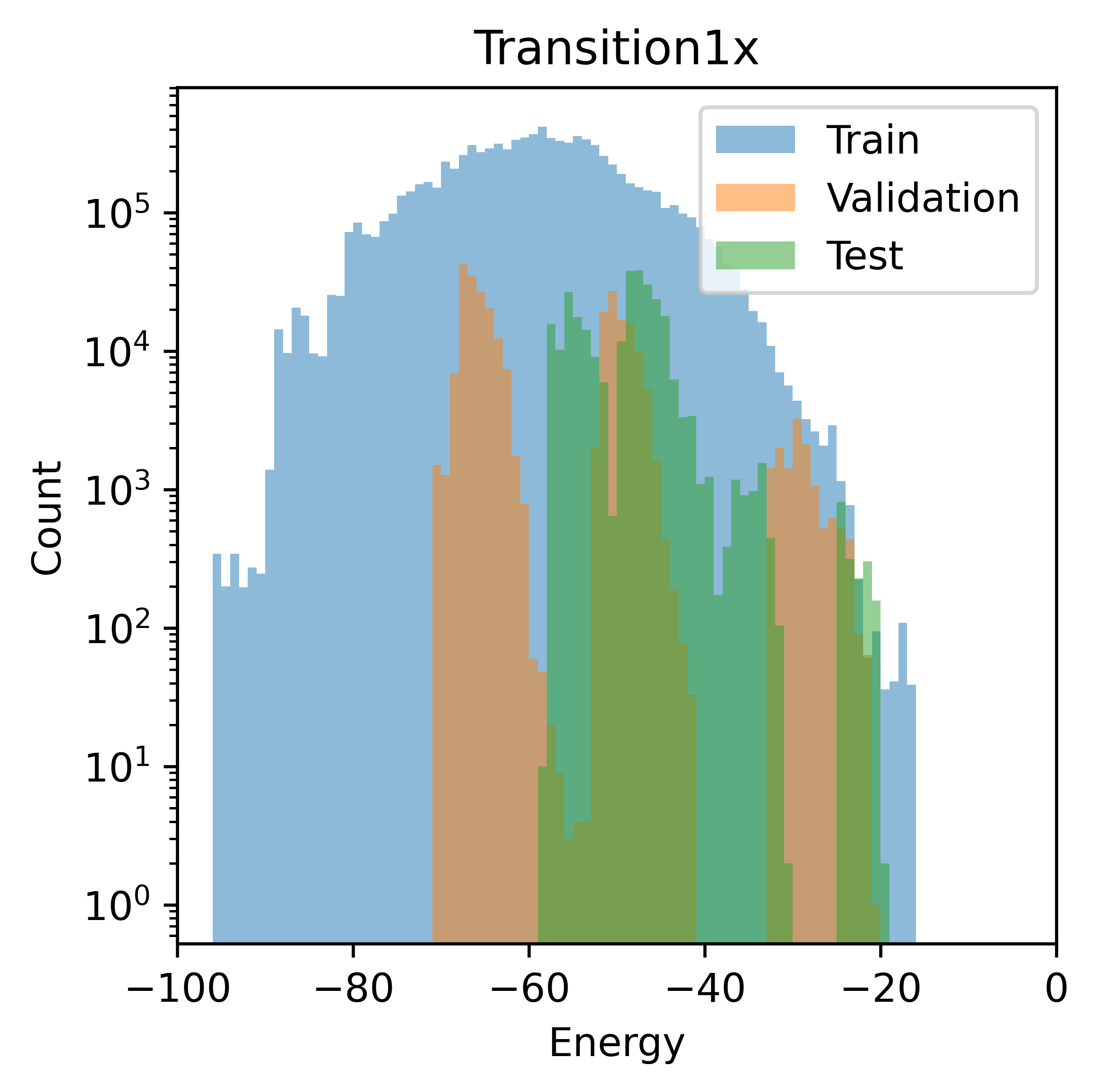}
\caption{Local root z-score variance (RZV) conditioned on observed energy on the Transition1x test dataset (top). Energy distribution in the Transition1x data split (bottom).}
\label{fig:t1x-energy-distribution}
\end{figure}

Test results for $M=5$ ensembles are presented in the last four rows of Table~\ref{tab:results}.
As in the first experiment, all ensembles achieved a low error on energy and forces in terms of MAE and RMSE compared to the results reported in~\citet{schreiner2022} (MAE=0.048 on energy and MAE=0.058 on forces) and training with NLL loss did not decrease performance in terms of prediction error in any case.
All four ensembles achieved acceptable average calibration in terms of NLL and RZV on the Transition1x test data.
Surprisingly, ensembles trained with MSE loss were as well or better calibrated than ensembles trained with NLL loss on this dataset.
All ensembles score similar high CV indicating uncertainty estimates are heteroscedastic and thus informative.
Calibration plots for an ensemble trained on Transition1x with NLL loss on energy and forces are presented in Figure~\ref{fig:calibration-t1x-nll-nll}.
The uncertainty vs.~error plots show that in general large errors are associated with large uncertainties as desired.
For both energy and forces some errors extend beyond 2-3 standard deviations of uncertainty indicating the error distributions have wider tails and may not be Gaussian in this case.
Similar to the ANI-1x experiment, it looks like the model is biased for some instances with large energy errors, but these cases are correctly identified as problematic by high uncertainty.
For the forces, the error distribution appears more symmetrical around zero but with wide tails.
The local z-score analysis plot for the energy indicate some inconsistencies in the energy uncertainties.
Plotting the root variance of the z-scores as a function of the observed molecular energies (Figure~\ref{fig:t1x-energy-distribution}) shows a tendency of the model to underestimate the uncertainty for low energies and overestimate the uncertainty for high energies on average.
This is a problem with the model that can not be corrected by scaling the uncertainties in the recalibration step.
Taking a closer look at the energy distribution reveals significant differences between the training, validation and test sets that is likely a consequence of splitting the data on chemical formula which could be the reason for this problem.
The variance of the local z-scores for the forces are more consistent, but values below one indicate that the model generally overestimates the uncertainty on the forces.
The reliability diagram for the energy also shows signs of some inconsistencies, as the curve does not form a straight line along the diagonal, but overall the uncertainties are correlated with the expected error.
The corresponding reliability diagram for the forces shows a more consistent result, only with a tiny overestimation of the force uncertainty. 
The reliability diagrams are summarised by ENCE scores in Table~\ref{tab:results}.
Additional calibration results are included in the ESI.

\section{Discussion}
\label{sec:discussion}

The proposed method achieved good predictive performance as well as calibrated and consistent uncertainty estimates in experiments on two challenging, publicly available molecular datasets.
A major advantage of the approach is that it considers both epistemic and aleatoric uncertainty through an ensemble approximation of mean-variance models. 
We believe that considering both aleatoric and epistemic uncertainty is critical to ensure good calibration in and out of the training data distribution.
Often the training procedures of uncertainty aware models do not inherently ensure good calibration on unseen data.
For example, ensemble members trained on the same data will often fit the same mean prediction without accounting for errors caused by random noise or inconsistency in the data or model inadequacy and mean-variance methods will estimate the expected error on the training data but do not guarantee good extrapolation of the uncertainty estimates to unseen data.
Therefore, the post hoc recalibration procedure is key to achieve good calibration on unseen data in our experiments, but is not commonly applied by other UQ methods in the literature.

The computational overhead of training and evaluating ensemble models is sometimes pointed out as a major disadvantage of using ensembles. 
However, it is important to note that most of this computation can be performed in parallel and thus only leads to a small overhead of computing the ensemble approximation and recalibration in real time.
Some works have proposed methods for speeding up the training of ensembles, such as snapshot ensembles~\cite{huang2017, wang2021}, which could also be applied in this case.
Another widely accepted advantage of ensembles is that they often improve prediction accuracy (see Figure~\ref{fig:ani1x-ensemble-size} as an example), which can be considered a positive side effect of the proposed method.
Here, we have used ensembles of size 5, but larger ensembles can be expected to further improve performance (up to a limit) at the cost of more computation.
The approach could also potentially benefit from other recent extensions to ensembles such as using randomized priors~\cite{randomizedprios} to improve the quality of especially the epistemic uncertainty estimates.

Evaluation of uncertainty calibration for regression models is an active area of research~\cite{kuleshov2018, levi2022, tran2020, pernot2022}.
Standard procedures for assessing the quality of uncertainty estimates are necessary within the field to establish confidence in individual UQ methods and ensure fair comparison. 
We recommend recent work by~\citet{pernot2023} which provides a good overview of calibration assessment methods and a detailed approach for evaluating uncertainty.
Our experiments show that the uncertainty estimates obtained with the proposed method are largely consistent with the expected error for varying size of the uncertainty (Figures~\ref{fig:calibration-ani1x-nll-nll} and~\ref{fig:calibration-t1x-nll-nll}).
However, we observed indications that uncertainties are not equally well calibrated along different molecular energies (Figure~\ref{fig:t1x-energy-distribution}). 
The current recalibration method only considers the magnitude of the predicted uncertainty. 
It would be an interesting direction for future work to design a recalibration function that can account for additional input features such as the (predicted) energy, while remaining a monotonic increasing scaling function, with the aim of achieving equally good calibration throughout the input space.
Applying the calibration evaluation framework proposed by~\citet{pernot2023} could help provide additional insights into the consistency and adaptivity of predictive uncertainty.

\section{Conclusion}
\label{sec:conclusion}

In this work, we have presented a complete framework for training neural network potentials with calibrated uncertainty estimates on both energy and forces.
The proposed method was demonstrated and evaluated on two challenging, publicly available molecular datasets containing diverse conformations far from equilibrium.
In all cases, the proposed method achieved low prediction error and good uncertainty calibration. 
On the ANI-1x dataset training with NLL loss improved the calibration over training with standard MSE loss.
On the Transition1x dataset, the same improvement was not observed and good calibration was achieved by training with standard MSE loss and applying post hoc nonlinear recalibration.
This could be because the validation and test data are more out of distribution in this case. 
The proposed method does not depend on the particular architecture of the neural network model, and can thus easily be adapted to new models in the future.
We hope that this work will contribute to better calibrated ML potentials and enable more robust and reliable applications.









\balance


\bibliography{references} 
\bibliographystyle{rsc} 

\clearpage
\appendix

\onecolumn
\setcounter{page}{1}
\counterwithin{figure}{section}

\section*{Electronic Supplementary Information:}


\subsection*{Graph Neural Network Interatomic Potential Ensembles\\with Calibrated Aleatoric and Epistemic Uncertainty\\on Energy and Forces}

\vspace{1em}

Jonas Busk, Mikkel N. Schmidt, Ole Winther, Tejs Vegge and Peter Bj{\o}rn J{\o}rgensen

\vspace{2em}

\section{Additional results}

Here we include additional calibration results from the experiments on the ANI-1x and Transition1x datasets presented in Section~\ref{sec:results} of the main paper.

Confidence curves can be used to evaluate the ranking ability of the model and to estimate the drop in error as a percentage of the high uncertainty instances are removed~\cite{soleimany2021, scalia2020}, which is especially important in applications such as active learning where high uncertainty instances are iteratively added to the training set to improve the model.
A confidence curve is generated by sorting predictions by uncertainty in decreasing order and computing the error as a function of removing a percentage of the most uncertain predictions.
For a well calibrated model the confidence curve is expected to decrease monotonically.
An oracle curve representing perfect ranking can be generated by sorting the prediction by error instead and the confidence curve can be summarised by computing the area between the confidence and oracle curves (AUCO).
However, we do not expect the uncertainty predictions to produce a perfect ranking with respect to the errors since instances with high predicted uncertainty can still have small empirical errors.

Quantile-calibration plots compare the quantiles of the predictive distribution with the quantiles of the empirical distribution and is a way to evaluate distribution calibration averaged over the data~\cite{kuleshov2018}.
If the predictive distribution matches the empirical distribution, the quantile-calibration curve corresponds to the identity function and forms a line along the diagonal of the plot.
Assuming a symmetric distribution, the confidence interval can be evaluated instead of the quantile. 
The quantile-calibration curve can be summarised by the sum of squared errors (SSE) between the predicted and empirical quantiles.

To further check the distribution assumptions averaged over the data, a histogram the errors normalised by the predicted uncertainties (z-scores) along with the assumed standard distribution can also be plotted.

\subsection{Additional ANI-1x results}

\begin{figure}[h]
\centering
\includegraphics[height=0.35\linewidth]{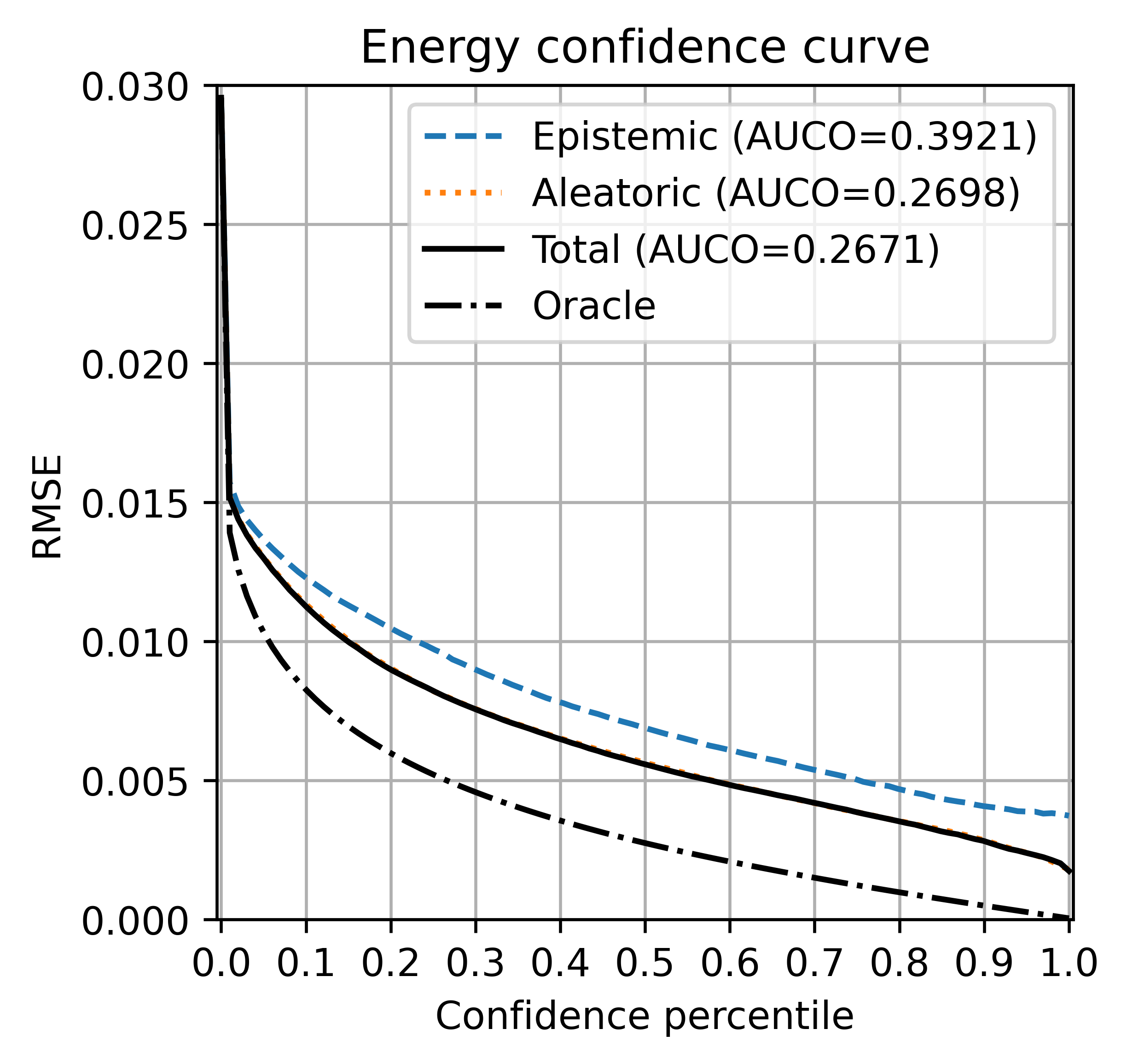}
\includegraphics[height=0.35\linewidth]{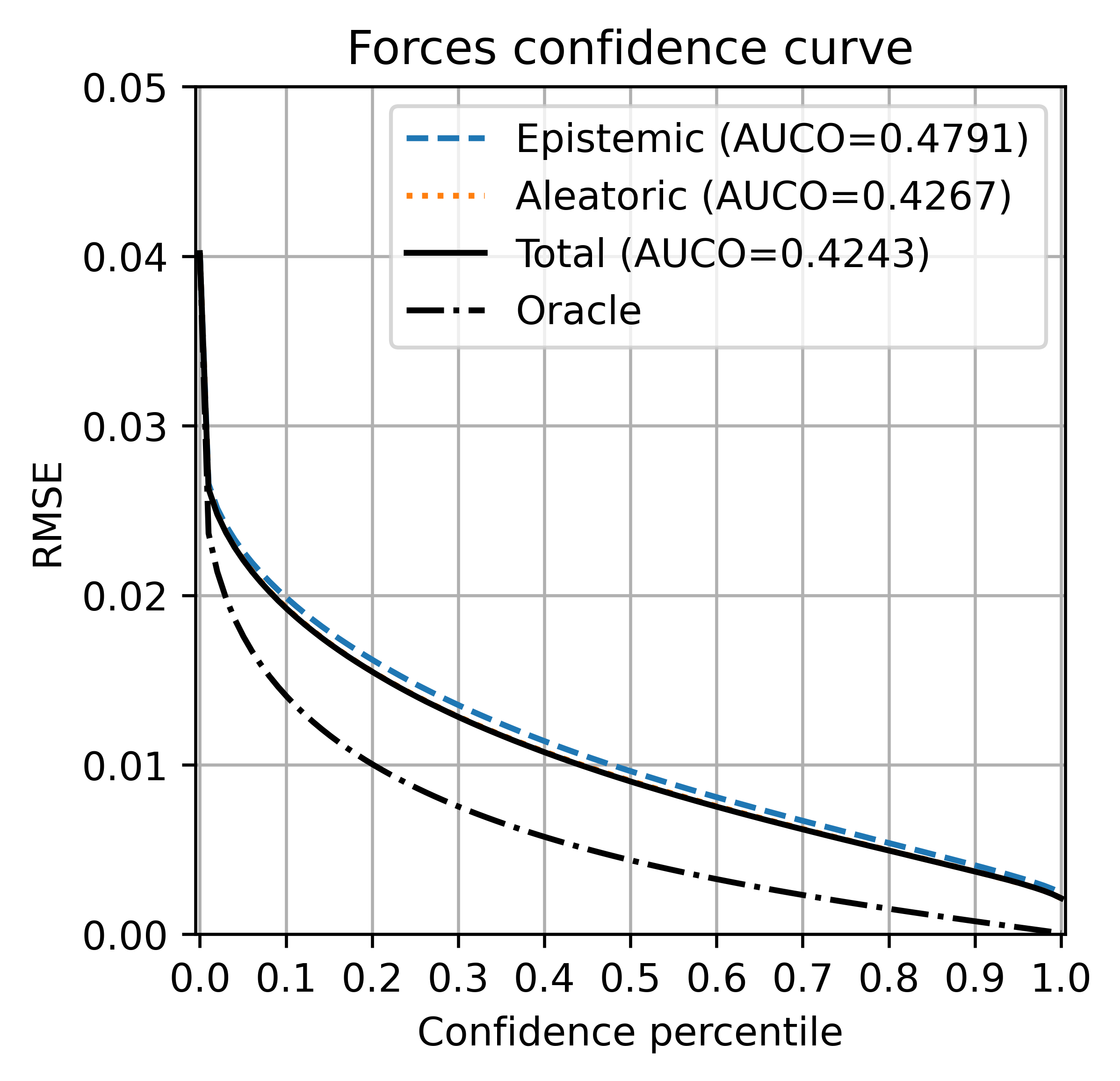}
\\
\includegraphics[height=0.35\linewidth]{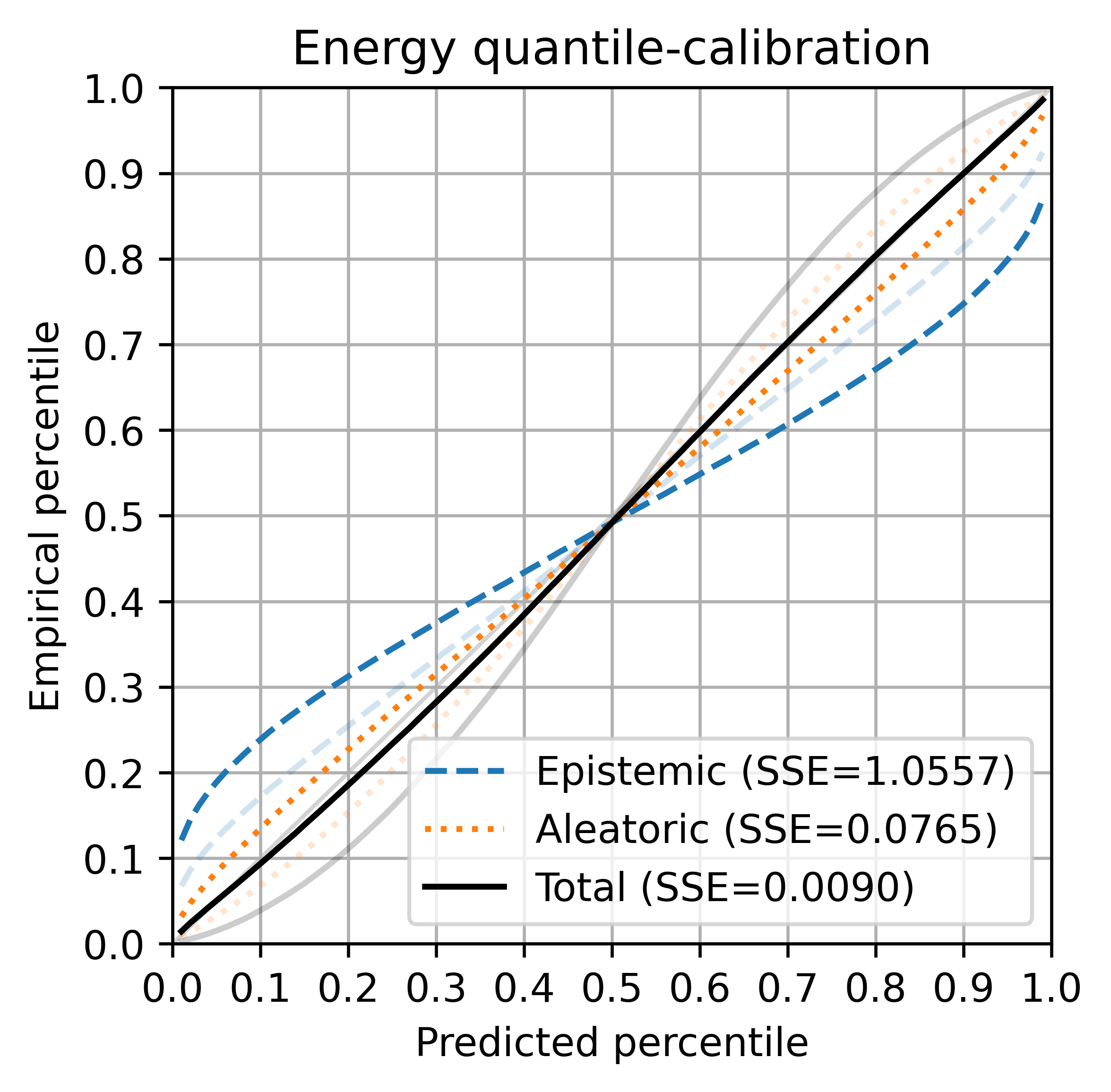}
\includegraphics[height=0.35\linewidth]{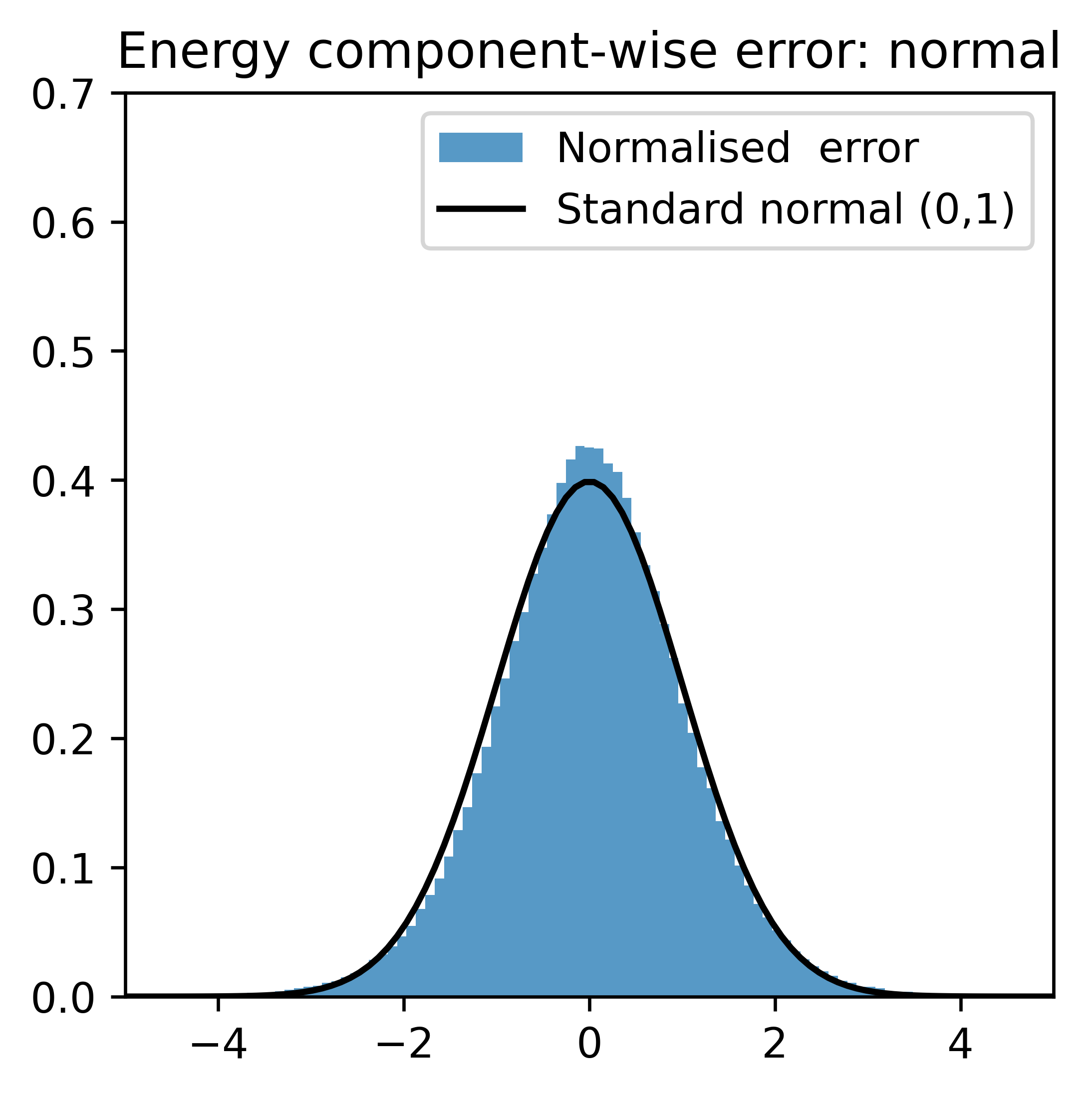}
\\
\includegraphics[height=0.35\linewidth]{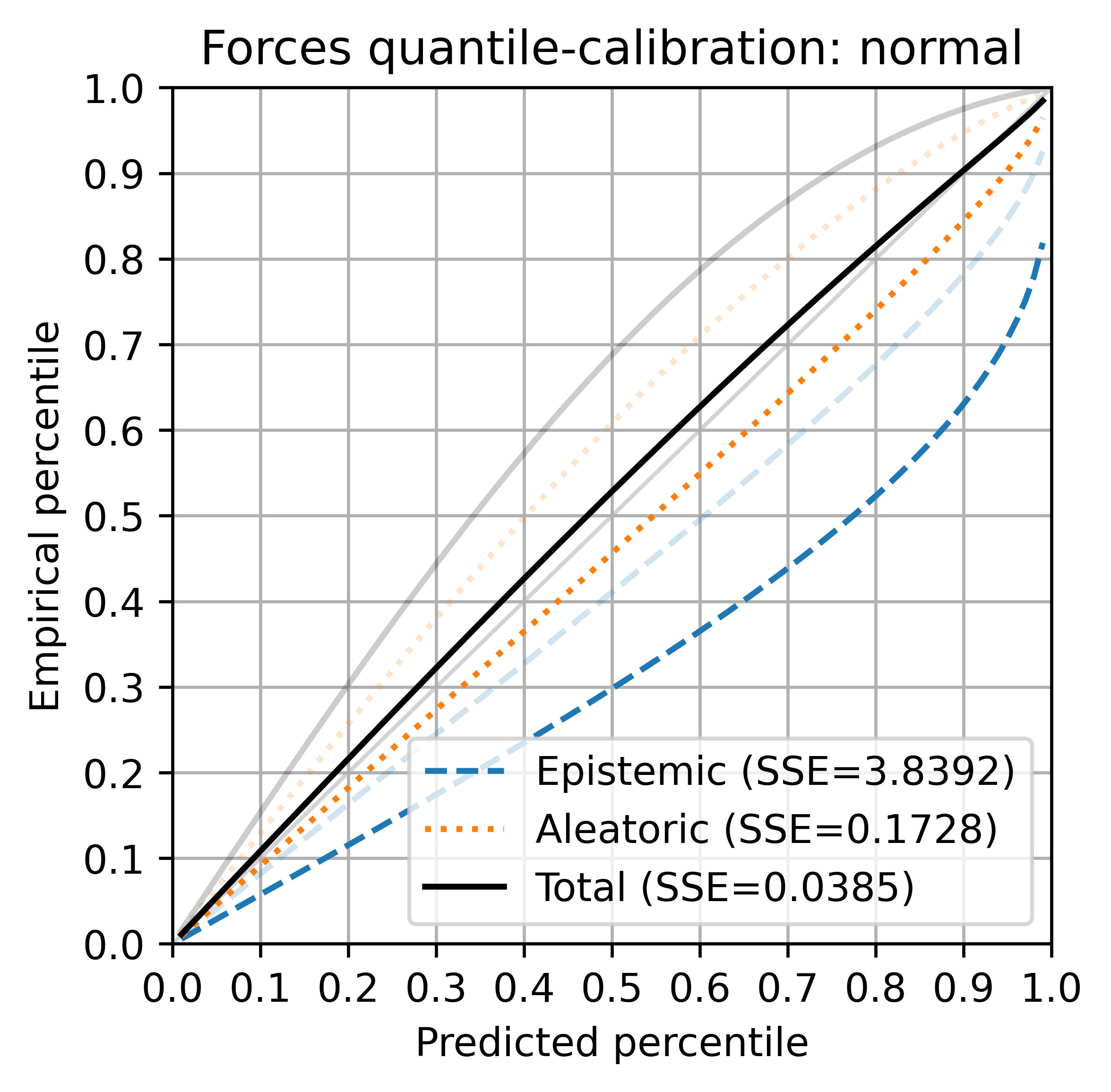}
\includegraphics[height=0.35\linewidth]{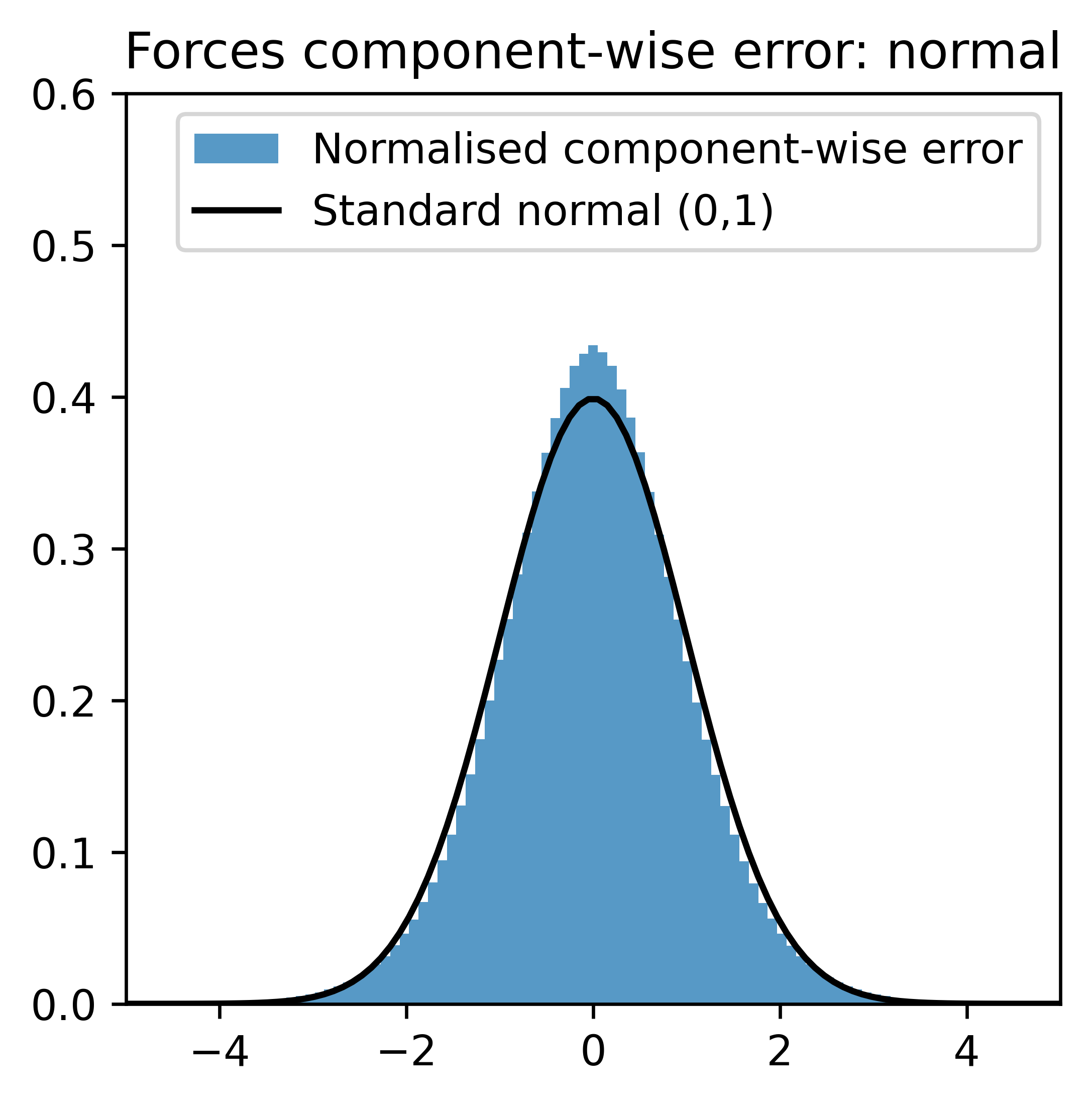}
\\
\includegraphics[height=0.35\linewidth]{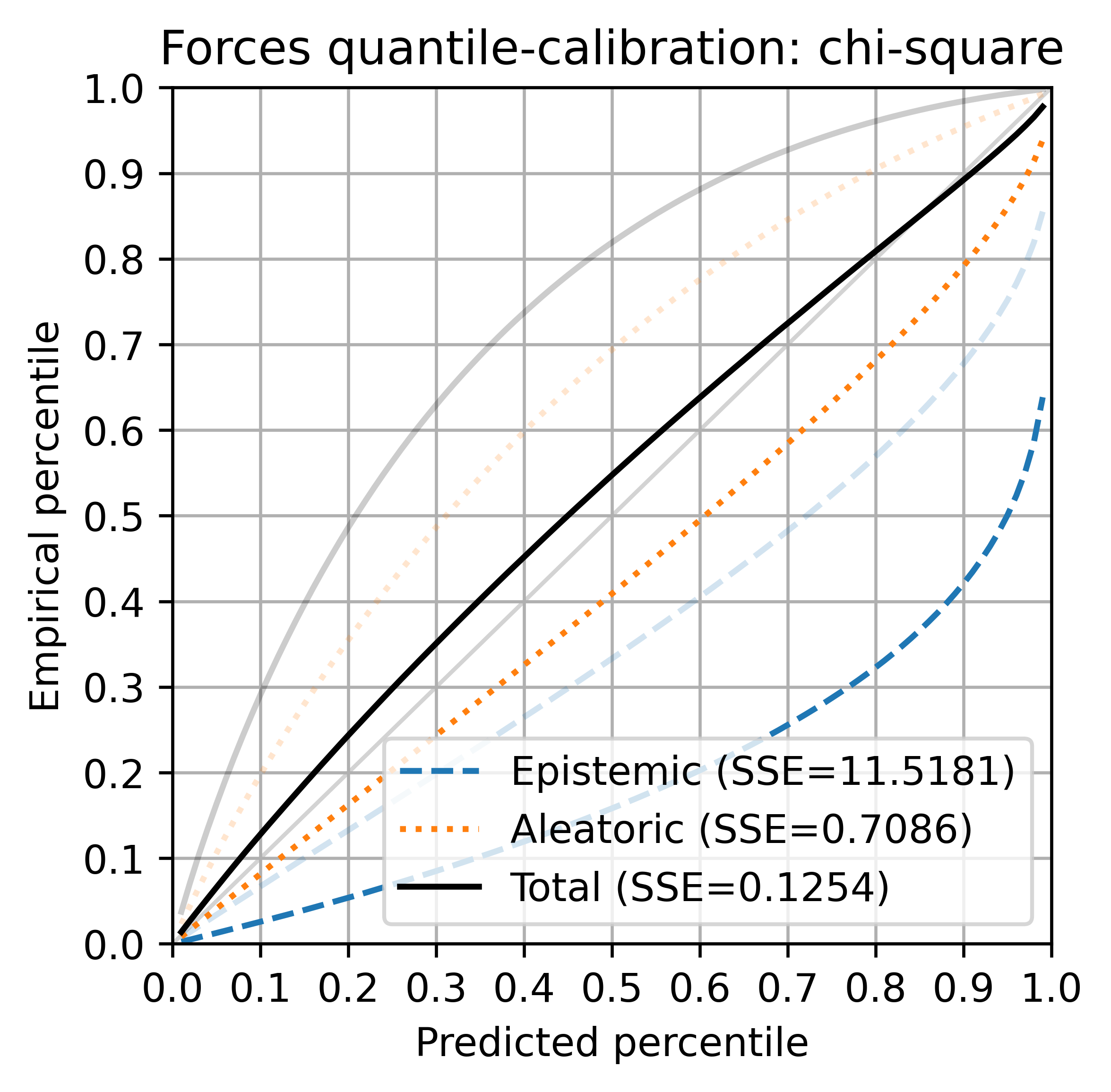}
\includegraphics[height=0.35\linewidth]{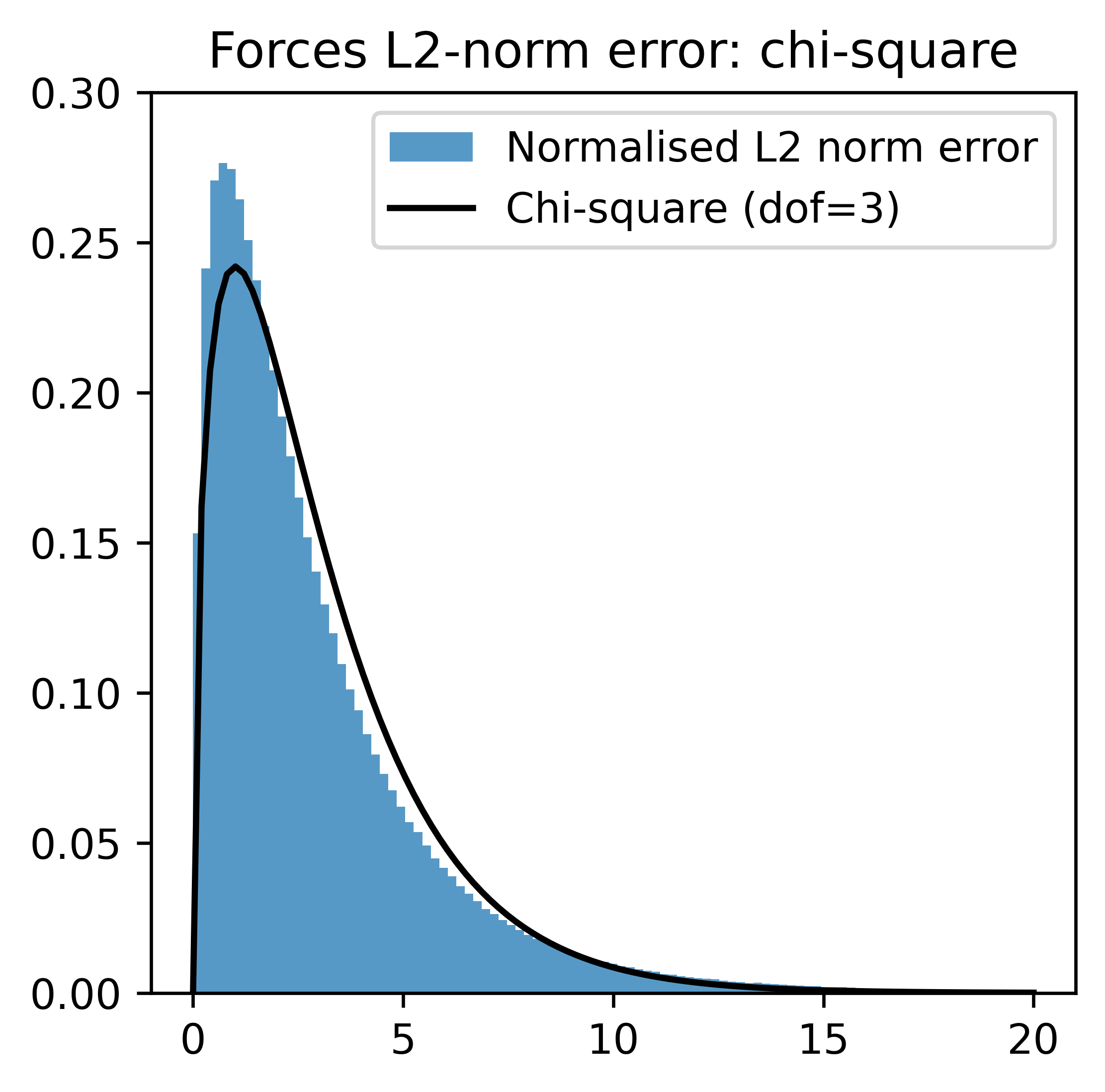}
\caption{Additional calibration results on the ANI-1x dataset of energy and forces for an ensemble of $M=5$ models trained with NLL loss on both energy and forces.
To illustrate the effect of recalibration, the transparent curves show results before applying recalibration whereas the solid curves show results after recalibration.}
\label{fig:additional-calibration-ani1x-nll-nll}
\end{figure}

Additional calibration plots for the ensemble model trained on ANI-1x with NLL loss on energy and forces are presented in Figure~\ref{fig:additional-calibration-ani1x-nll-nll}.
The confidence curves for energy and forces are both monotonically decreasing indicating good ranking ability. 
The confidence curves also show a significant drop in error on both energy and forces when removing the top $\sim5\%$ highest uncertainty predictions.
This confirms the observation from the reliability diagrams in Figure~\ref{fig:calibration-ani1x-nll-nll}, that there are a few instances with very high error but they are correctly identified and assigned high uncertainty by the model.

The energy quantile-calibration plot shows that assuming a normal distribution percentiles of the predicted distributions corresponds well to the empirical distribution and the symmetry at the 0.5 percentile indicates that the model is unbiased overall.
In the case of the forces, we assume the distribution of the component-wise errors is normal and unbiased.
Furthermore, if the component-wise force errors are normally distributed, the squared L2 norm of the 3-dimensional force errors should follow a chi-square distribution with 3 degrees of freedom.
Consequently, we plot the symmetric version of the quantile-calibration plot for the component-wise force errors using a normal distribution and a regular quantile plot for the squared L2 norm of the force errors using a chi-square distribution as they allow for easier comparison.
In both cases, the uncertainty estimates look fairly well calibrated with regards to the assumed distributions.
This is also apparent from the corresponding histograms of normalised errors plotted along with the reference distributions.

\subsection{Additional Transition1x results}

\begin{figure}[h]
\centering
\includegraphics[height=0.35\linewidth]{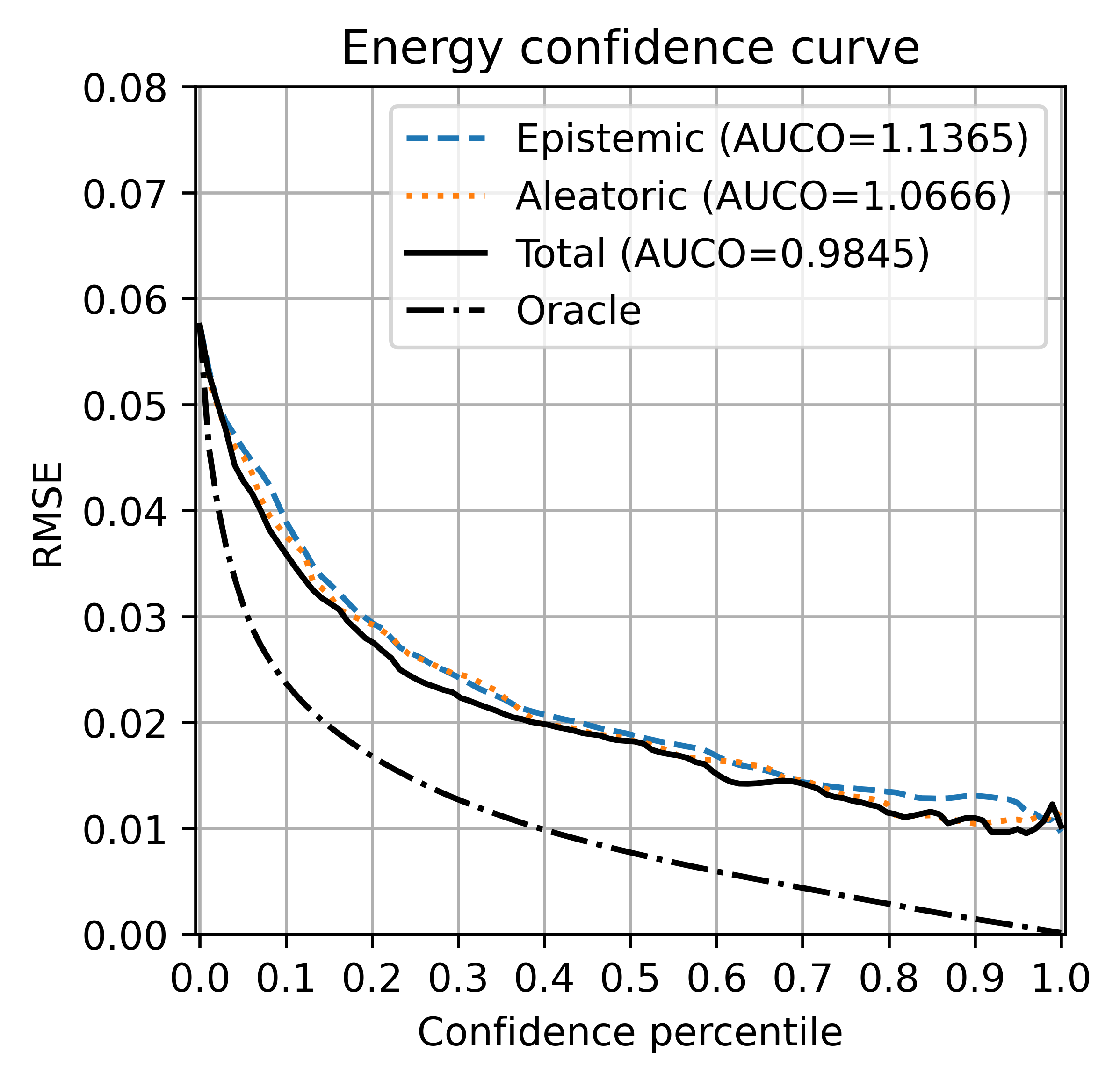}
\includegraphics[height=0.35\linewidth]{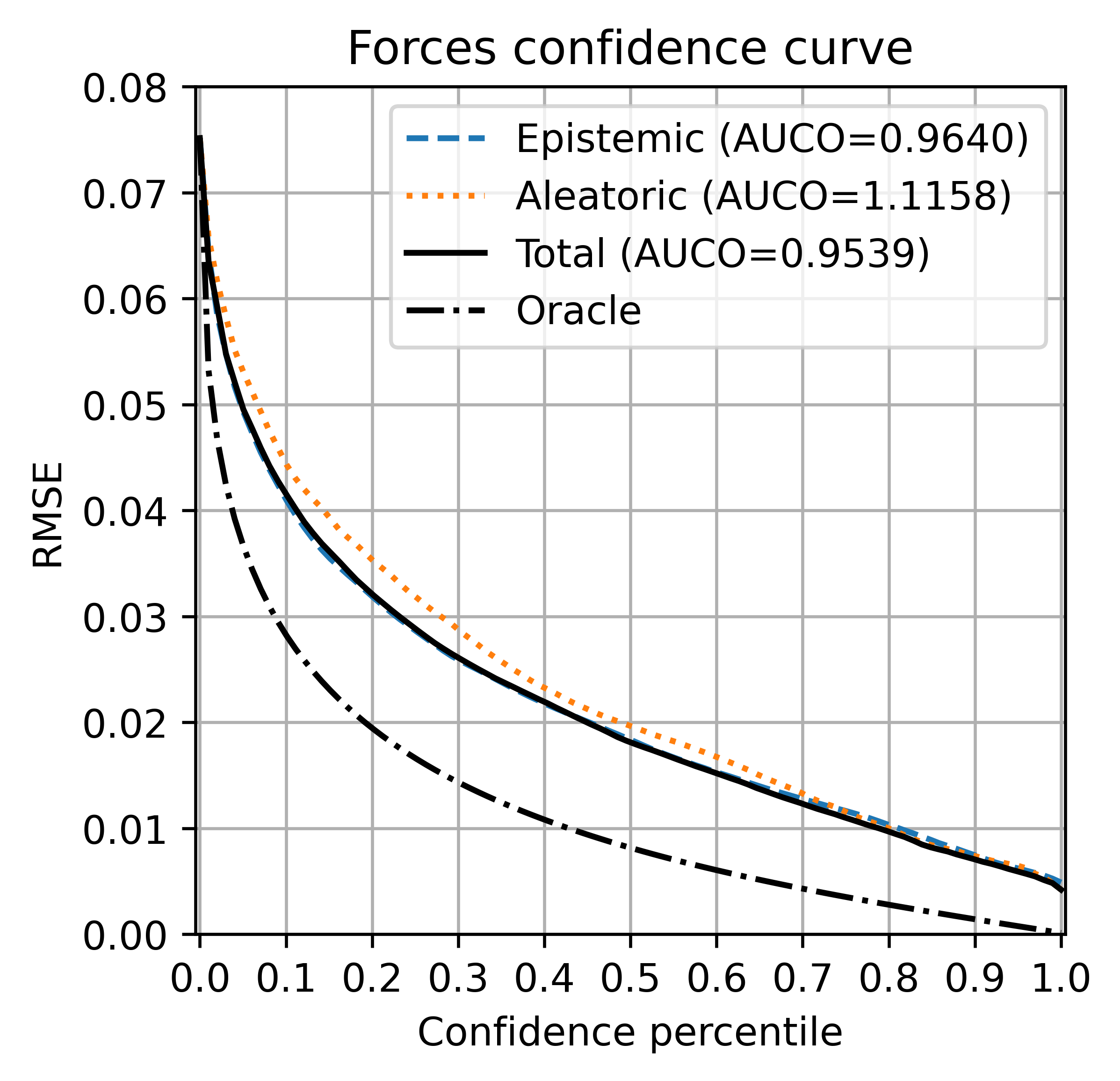}
\\
\includegraphics[height=0.35\linewidth]{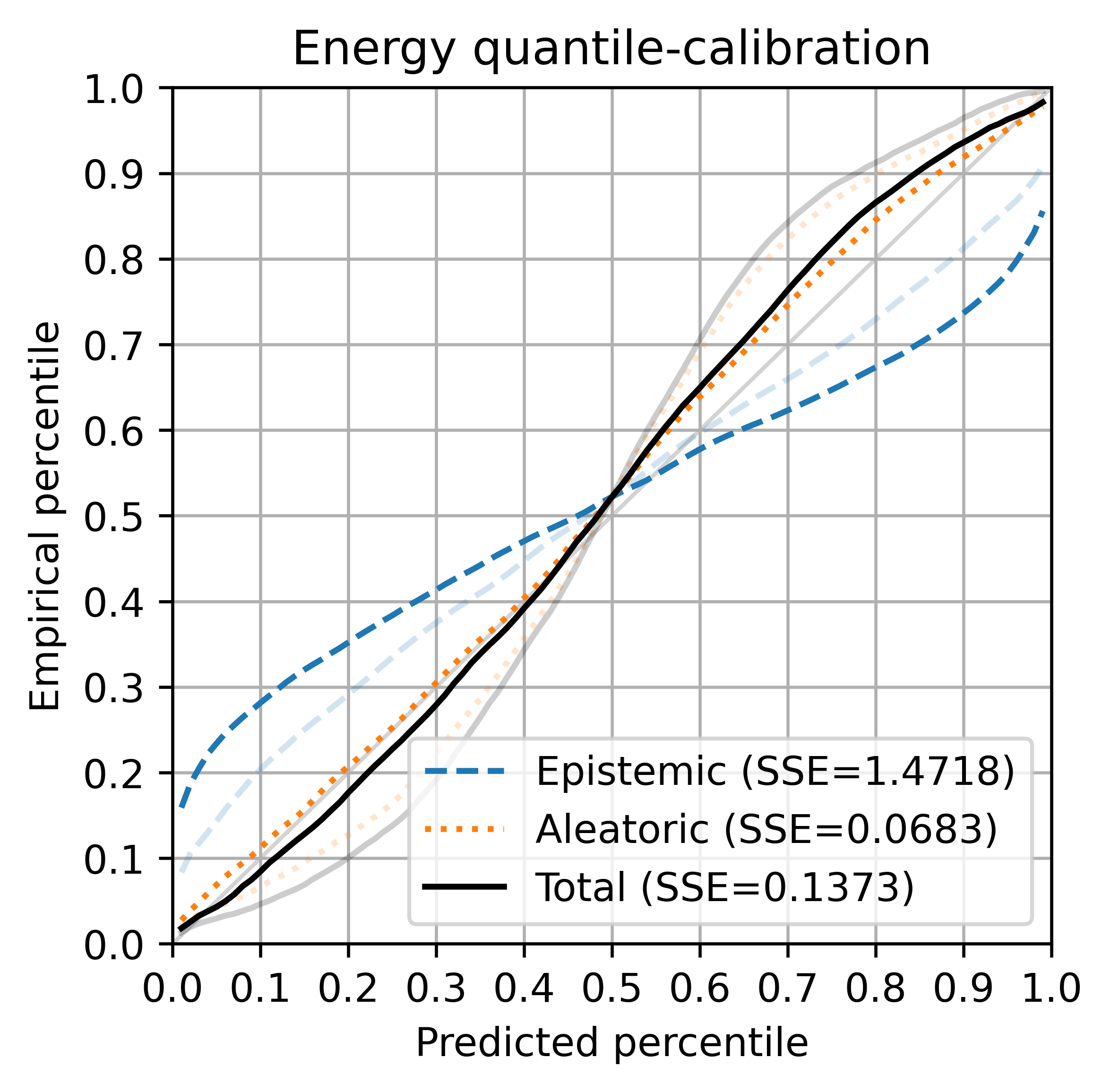}
\includegraphics[height=0.35\linewidth]{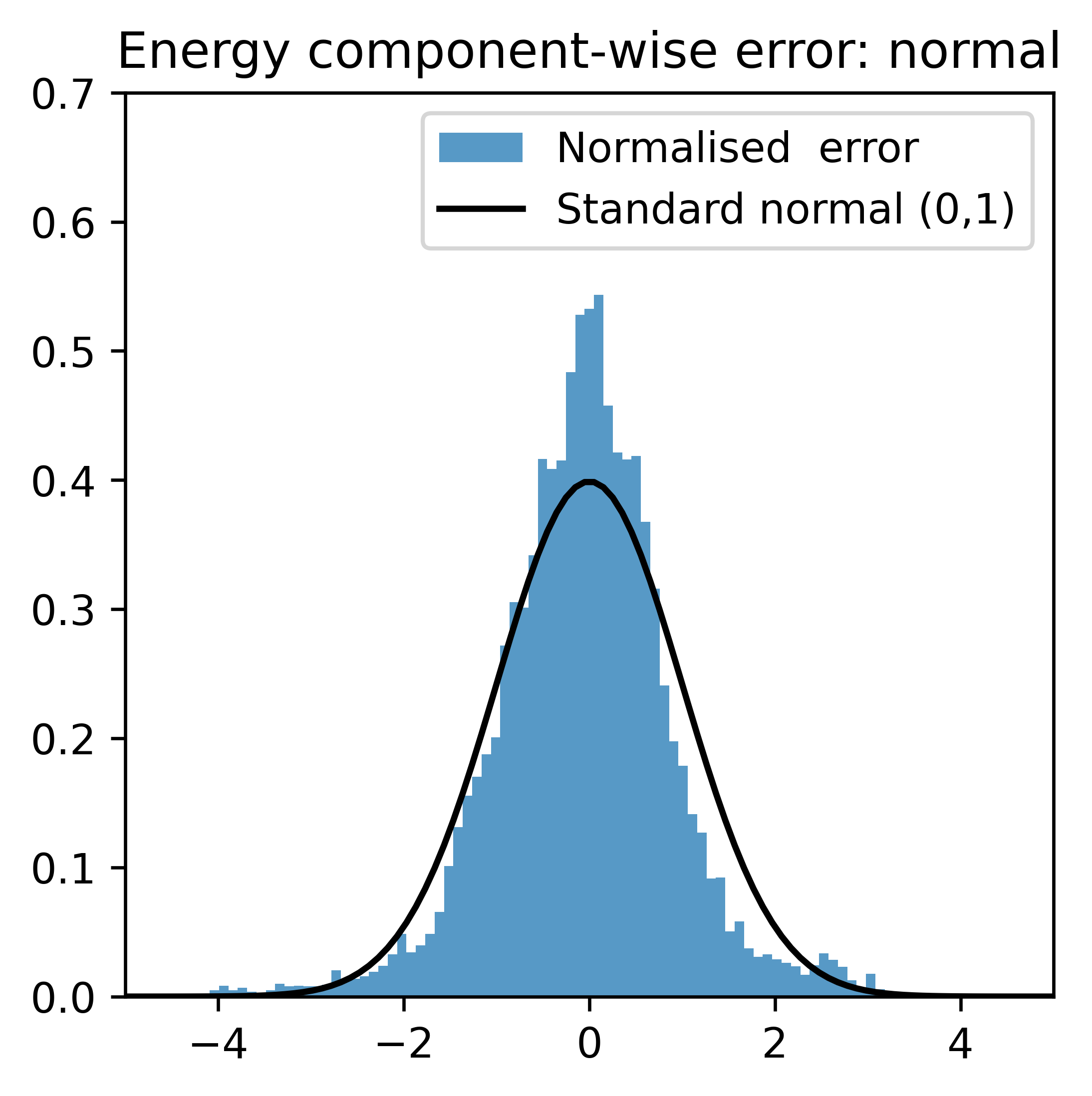}
\\
\includegraphics[height=0.35\linewidth]{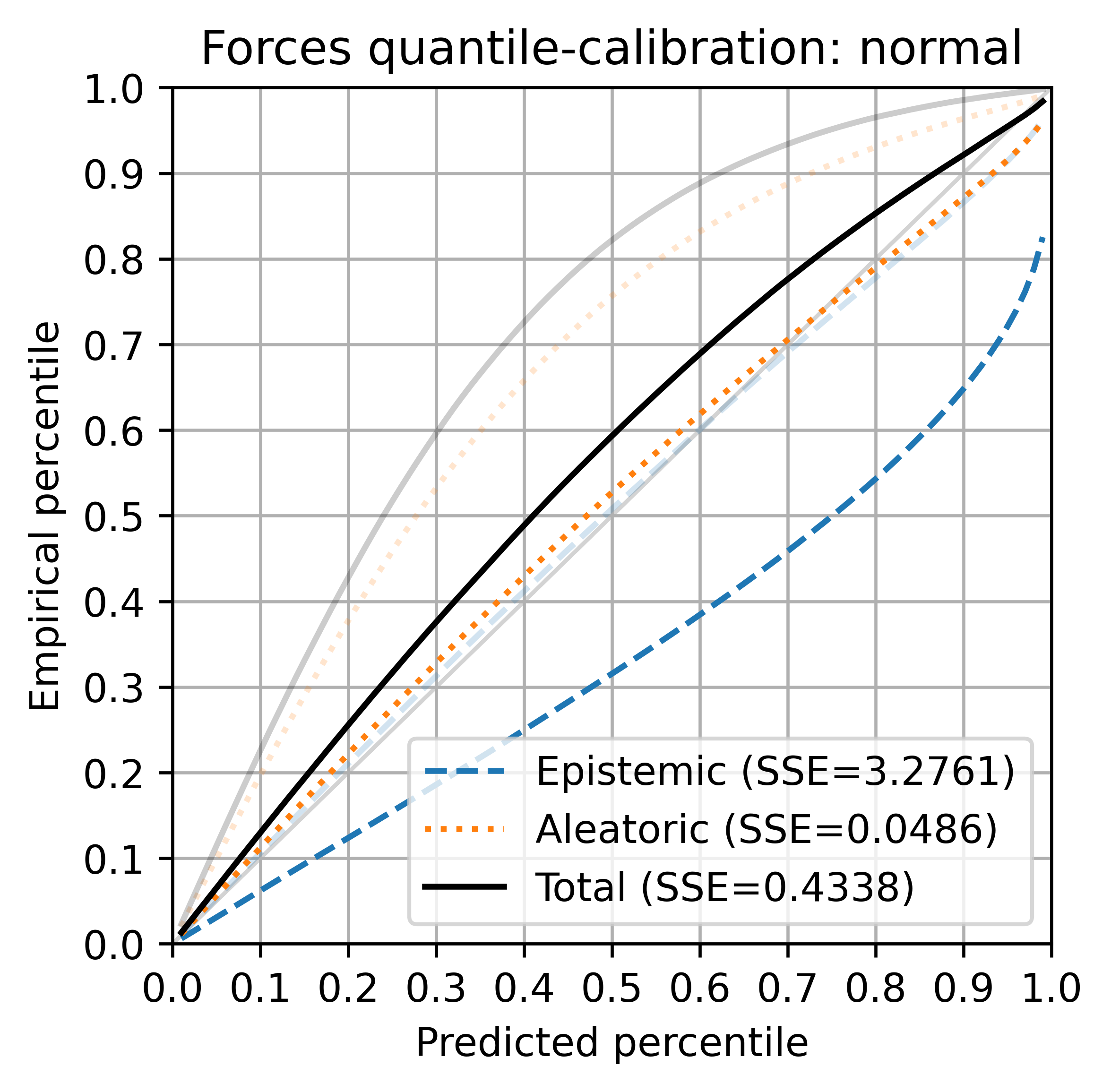}
\includegraphics[height=0.35\linewidth]{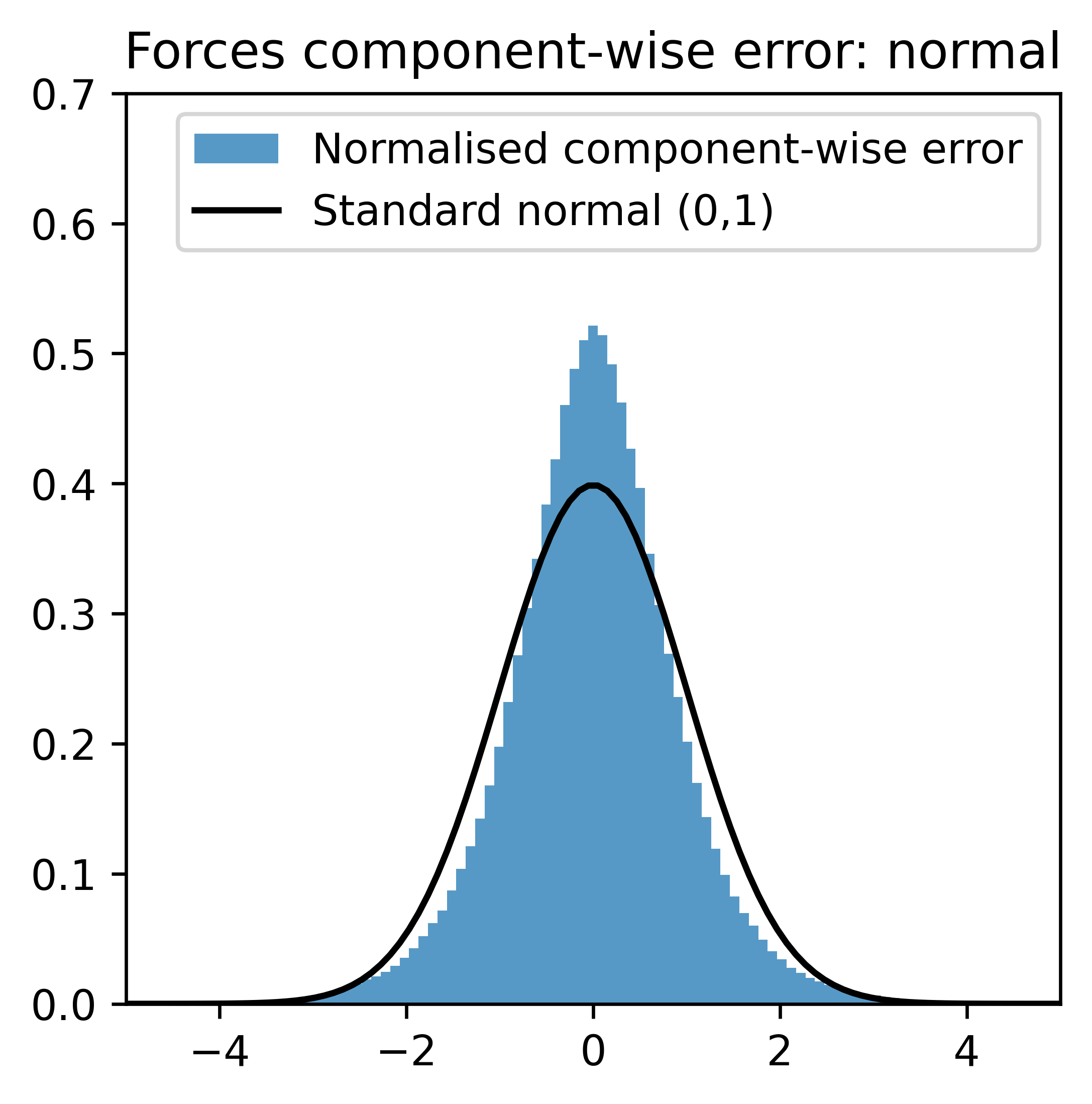}
\\
\includegraphics[height=0.35\linewidth]{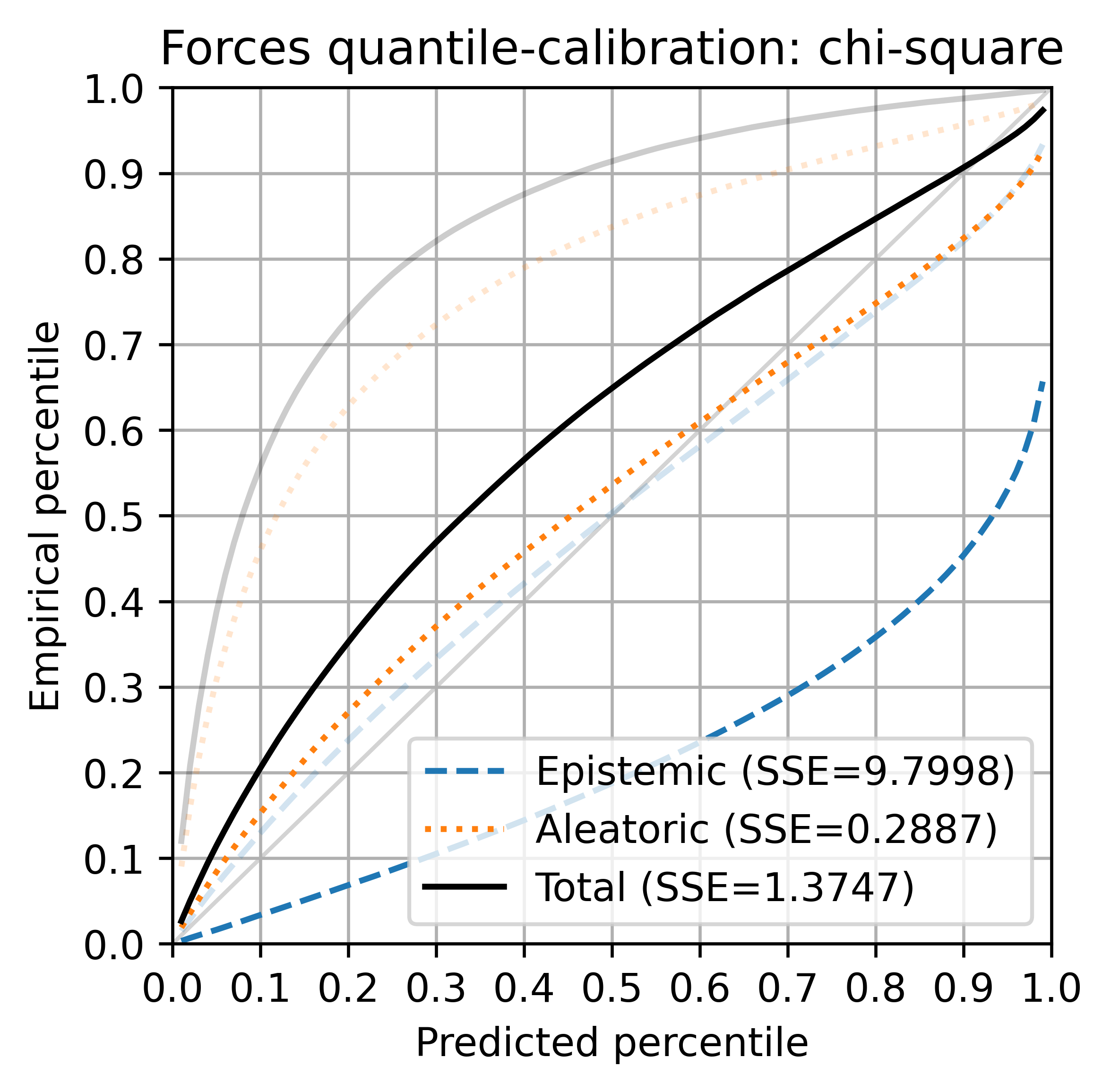}
\includegraphics[height=0.35\linewidth]{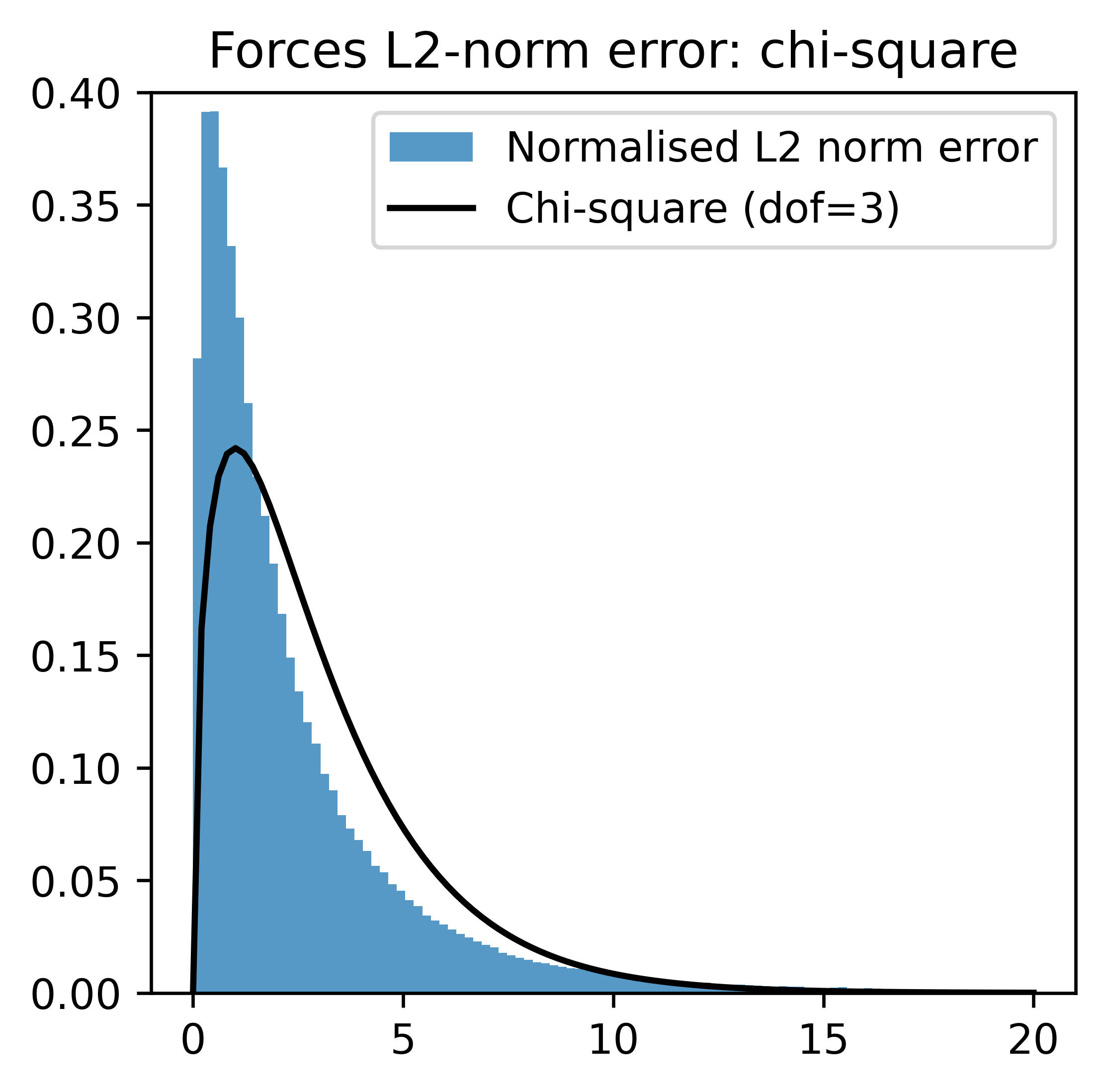}
\caption{Additional calibration results on the Transition1x dataset of energy and forces for an ensemble of $M=5$ models trained with NLL loss on both energy and forces.
To illustrate the effect of recalibration, the transparent curves show results before applying recalibration whereas the solid curves show results after recalibration.}
\label{fig:additional-calibration-t1x-nll-nll}
\end{figure}

Additional calibration plots for the ensemble model trained on Transition1x with NLL loss on energy and forces are presented in Figure~\ref{fig:additional-calibration-t1x-nll-nll}.
The energy confidence curve is generally decreasing, but like the corresponding reliability diagram shown in Figure~\ref{fig:calibration-t1x-nll-nll} it is not perfectly consistent whereas the forces confidence curve look more consistent and monotonically decreasing.
In both cases, removing the highest uncertainty instances results in a large drop en error.

The energy quantile-calibration plot shows that the error is fairly well distribution calibrated with a slight overestimation of the uncertainty on average.
The same applies to the forces where the error is also fairly well distribution calibrated but with some overestimation of the uncertainty.
This is consistent with the results of the LZV analysis described in Section~\ref{sec:t1x-results}.

\end{document}